\documentclass[aps,prd,superscriptaddress,twocolumn,lengthcheck,superscriptaddress,nopreprintnumbers,nofootinbib]{revtex4-1}

\usepackage[utf8]{inputenc}
\usepackage{aas_macros}
\usepackage{natbib}

\usepackage{amsmath,amssymb,amsfonts}
\usepackage{bbm}
\usepackage{graphicx} 
\usepackage{rotating}
\usepackage[caption=false]{subfig}
\usepackage{bm}
\usepackage{tabularx}
\usepackage[usenames]{color}
\usepackage{comment}

\usepackage{xspace}
\usepackage[plainpages=false, colorlinks=true, anchorcolor=blue, linkcolor=cyan, citecolor=cyan, bookmarks=false]{hyperref}

\newcolumntype{b}{X}
\newcolumntype{s}{>{\hsize=.5\hsize}X}

\def\be{\begin{equation}}
\def\ee{\end{equation}}
\newcommand{\bb}{\begin{bmatrix}}
\newcommand{\eb}{\end{bmatrix}}

\def\bea{\begin{align}}
\def\eea{\end{align}}

\def\be{\begin{equation}}
\def\en{\end{equation}}
\def\bea{\begin{eqnarray}}
\def\ena{\end{eqnarray}}

\begin{document}

\title{From Bright Binaries To Bumpy Backgrounds: \\Mapping Realistic Gravitational Wave Skies With Pulsar-Timing Arrays}

\author{Stephen~R.~Taylor}
\email[]{stephen.r.taylor@vanderbilt.edu}
\affiliation{Department of Physics \& Astronomy, Vanderbilt University, 2301 Vanderbilt Place, Nashville, TN 37235, USA}
\affiliation{TAPIR Group, California Institute of Technology, 1200 East California Boulevard, Pasadena, CA 91125, USA}

\author{Rutger~van~Haasteren}
\affiliation{TAPIR Group, California Institute of Technology, 1200 East California Boulevard, Pasadena, CA 91125, USA}
\affiliation{Microsoft Corporation, One Microsoft Way, Redmond, WA 98052, USA}

\author{Alberto~Sesana}
\affiliation{Università degli Studi di Milano-Bicocca, Piazza della Scienza 3, 20126 Milano, Italy}

\date{\today}

\begin{abstract}
Within the next several years, pulsar-timing array programs will likely usher in the next era of gravitational-wave astronomy through the detection of a stochastic background of nanohertz-frequency gravitational waves, originating from a cosmological population of inspiraling supermassive binary black holes. While the source positions will likely be isotropic to a good approximation, the gravitational-wave angular power distribution will be anisotropic, with the most massive and/or nearby binaries producing signals that may resound above the background. We study such a realistic angular power distribution, developing fast and accurate sky-mapping strategies to localize pixels and extended regions of excess power while simultaneously modeling the background signal from the less massive and more distant ensemble. We find that power anisotropy will be challenging to discriminate from isotropy for realistic gravitational-wave skies, requiring SNR $>10$ in order to favor anisotropy with $10:1$ posterior odds in our case study. Amongst our techniques, modeling the population signal with multiple point sources in addition to an isotropic background provides the most physically-motivated and easily interpreted maps, while spherical-harmonic modeling of the square-root power distribution, $P(\hat\Omega)^{1/2}$, performs best in discriminating from overall isotropy. Our techniques are modular and easily incorporated into existing pulsar-timing array analysis pipelines.
\end{abstract}


\maketitle

\section{Introduction}
\label{sec:intro}

High-precision timing of millisecond pulsars (MSPs) has been, and continues to be, a valuable tool for probing a wide range of physics, from studies of nuclear matter to tests of modified gravity theories. Successes include the first indirect confirmation of gravitational-wave (GW) emission \cite{tw82,twdw92}, and very accurate tests of general relativity \cite{ksm+2016,2018Natur.559...73A}. Pulsar-timing array (PTA) projects aim to \emph{directly} detect low-frequency GWs in the range $10^{-9}$--$10^{-7}$~Hz from extra-Galactic sources by using a set of Galactic MSPs as nearly-perfect Einstein clocks \citep{fb90}. This endeavor is possible due to the exceptional regularity of pulses and the remarkable stability of pulse profiles. Many processes must be modeled in a pulsar's timing ephemeris, but if done so accurately one can account for every rotation of the pulsar across observation epochs. The presence of GWs affects the propagation of pulses from the pulsar to the Earth, changing the proper length of the photon path, and creating detectable deviations away from the expected pulse times of arrival (TOAs)  \citep{ew75,saz78,det79}.

Over the past two decades, several regional collaborations have been collecting data and hunting for a stochastic GW background, periodic GW signals, GW memory bursts, and more; these include the European Pulsar Timing Array (EPTA, \citep{kc13}) \cite{vhj+11,ltm+15,bps+15},
the North American Nanohertz Observatory for Gravitational Waves (NANOGrav, \citep{ml13}) \cite{dfg+12,abb+14,abb15_bwm,abb+16,abb+18b,aab+19,aab20_bwm}, and the Parkes Pulsar Timing Array (PPTA, \citep{mhb+13,h13}) \citep{src+13,zhw+14,whc+15,srl+15,2016MNRAS.455.3662M}. These three were founding members of the International Pulsar Timing Array (IPTA, \citep{vlh+16,pdd+19}), aimed at strengthening pulsar sky coverage and observation baselines. There are also burgeoning PTA GW search efforts in India (InPTA, \citep{inpta}), China (CPTA, \citep{cpta}), as well as in telescope-centered timing groups like MeerTime \citep{meertime} and CHIME/Pulsar \citep{chime}. These ongoing collaborative searches have yielded precision timing data on more than $80$ MSPs, some with baselines longer than $15$ years. 

Binary systems of supermassive black holes (SMBHs) with masses $\sim 10^8-10^{10}M_\odot$ are expected to be the dominant source of GWs for PTAs \citep{bbr80,wl03,jb03,svc08}. Such titanic black holes lie at the heart of massive galaxies, forming binary systems as their host galaxies merge and grow over the history of the Universe \citep{kormendy95,magorrian98,fm00,gebhardt00}. After the merger of two galaxies, the resident SMBHs will each experience dynamical friction in the merger remnant, carrying them to $\sim 1-10$ parsec separations to form a bound supermassive binary black hole (SMBBH) system. At this stage, discrete scattering events with central bulge stars will evolve the binary toward even closer separations, possibly giving way to viscous interactions with a circumbinary disk until the binary becomes dominated by GW emission at separations of $\lesssim 10^{-3}$ parsecs \citep[see e.g.,][and references therein]{sbs19}. The system will then be emitting GWs at frequencies that are accessible to PTAs. While PTAs are in principle sensitive to the entire cosmological population of SMBBHs, discriminating each and every one is beyond the resolution limit \citep{bs12,pbs+12,bp12}. Hence the first expected signal will be that of the incoherent superposition of all binaries producing a stochastic GW background (GWB) \citep{rsg15,kbh+18,s13}, above which a few particularly nearby or massive binaries may resound as individually detectable signals \citep{svv09,sv10,lwk+11,bs12,esc12,ellis13,teg14,zhw+14,thgm16}.

Beyond a few hundred Mpc, the Universe can be approximated as statistically homogeneous and isotropic. Hence the (an)isotropy of the GW signal across the sky is highly dependent on the underlying demographics and finiteness of the SMBBH population, rather than the actual positional distribution of sources. Current estimates place deviations from isotropy in the nanohertz GW sky below $\sim 10\%$ \citep{msmv13,tg13,mls+17,roebber17}, making anisotropy a challenging yet rewarding scientific goal after the initial isotropic GWB detection by PTAs. The assumption of GWB isotropy leads to a simple analytic expression for the induced inter-pulsar timing correlations that is only dependent on the angular separation of the pulsars; this is known as the Hellings \& Downs curve \citep{hd83}. Correlations from anisotropic signatures can also be expressed analytically through a spherical harmonic decomposition of the angular power \citep{2009PhRvD..80l2002T,msmv13,tg13,grt+14}. Current search strategies for anisotropy implement a prior condition that blocks off unphysical regions of the spherical harmonic coefficients that would otherwise imply negative angular power \citep{tg13,tmg+15}. This prior reduces the efficiency of posterior sampling, and prevents gradient-based sampling methods from being used. More general methods that map the polarization content of the GWB using CMB techniques \citep{grt+14,cjm19,hkj19}, or decompose the angular power distribution on eigen-skies of the PTA response map \citep{cvh14,2020arXiv200614570A}, are also possible but have not yet been implemented within realistic analysis studies. 

The techniques developed in this paper represent a major step forward in mapping realistic gravitational-wave skies with PTAs (beyond those in \citet{tmg+15}), corresponding to methods that can be easily embedded in modern PTA pipelines, and that better account for anisotropy arising from bright single sources. Anisotropy searches should be superseded by joint searches for a stochastic GW background plus individual binary signals that can be resolved out of the binary ensemble. These kinds of searches will employ trans-dimensional search strategies that use the data to not only constrain the properties, but also the number, of favored sources \citep{pbs+12,bs12,bc19}. Sky localization of these individual binaries will initially be quite poor ($\sim\mathcal{O}(100)$ deg$^2$) \citep{sv10,thgm16,gsh+19}, necessitating known electromagnetic-counterpart candidates \citep{graham+15,jll+04,dhd+15,cbh+16,lga+19,hss+18,shk+18,zt20}, or priors on likely host-galaxy properties to permit finer resolution \citep{spl+14,mls+17,gsh+19}.

After the initial detection of an isotropic GWB, and prior to the resolution of individual bright binaries within this background, these binaries will announce themselves merely as regions of excess angular power. Anisotropy searches only constrain the incoherent angular power distribution, so there is no Occam penalty for the many binary model parameters that otherwise appear in individual source searches; the drawback of course is that no binary characterization is possible. Probes of anisotropy form an important link in the chain of scientific milestones for PTAs, bridging the information gap between background detection and individual source resolution. We study this link here, developing fast and flexible mapping strategies for the angular power distribution of the GWB, and studying which of these is most apt for the GW sky from a SMBBH population. We employ Bayesian model selection to make data-driven decisions on the number of excess power regions that can be isolated, and forecast the signal-to-noise ratio at which anisotropy will be favored over isotropy. 

This paper is laid out as follows. In Section \ref{sec:anisotropy} we introduce the formalism for computing the overlap reduction function between arrays of pulsars in the presence of an anisotropic stochastic GW background. We also discuss techniques to rapidly compute this under generic map parametrizations. In Sec.~\ref{sec:likelihood} we introduce our PTA likelihood for fitting simulated data to models, which is a simplified version of the full production-level model used in real PTA GW searches. In Sec.~\ref{sec:simulations} we discuss the two different spatial distributions of GW power that we consider here, corresponding to a region of excess power (a ``hotspot'') and a map composed of many individual SMBBH signals drawn from a population synthesis simulation. We present our results in Sec.~\ref{sec:results}, followed by concluding remarks in Sec.~\ref{sec:conclusions}, and supplementary calculations in the appendices.

\section{Anisotropy}
\label{sec:anisotropy}

In PTA searches for a Gaussian, stationary stochastic GW background, the primary detection statistic is one that leverages two-point inter-pulsar correlated timing deviations. At the center of this statistic lies the overlap reduction function (ORF), which describes the degree with which GW-induced timing deviations are correlated between Earth-pulsar systems that vary in orientation with respect to angular power on the sky. The pieces needed to compute the ORF include the antenna response function of each Earth-pulsar system, along with a representation of the underlying angular distribution of GW power. It is the latter dependence that allows us to probe angular structure in the nanohertz GW sky, in a bid to search for bright sources, clustered power regions, or statistical isotropy. In the following we describe different parametrized representations of this power, and test these with simulations to understand the most apt and statistically parsimonious representation of a given GW sky.

\subsection{Antenna Response} \label{sec:antenna}

A GW in General Relativity can be written as a sum over its two tensor transverse polarizations:
\begin{equation}
        h_{\mu\nu}(t,\hat{k}) = e^{+}_{\mu\nu}(\hat{k})h_{+}(t,\hat{k}) +
        e^{\times}_{\mu\nu}(\hat{k})h_{\times}(t,\hat{k}),
        \label{sec:metpert}
    \end{equation}
where $\hat{k}$ is the direction of GW propagation, $h_{+,\times}$ are the polarization amplitudes of the metric perturbation, and $e^{+,\times}_{\mu\nu}$  are the polarization basis tensors. The polarization basis tensors are constructed from a basis triad, such that:
\begin{equation} \label{eq:poltensors}
    e^{+}_{\mu\nu}(\hat{k}) = \hat{m}_{\mu}\hat{m}_{\nu} -
    \hat{n}_{\mu}\hat{n}_{\nu}, \quad e^{\times}_{\mu\nu}(\hat{k}) = \hat{m}_{\mu}\hat{n}_{\nu} +
    \hat{n}_{\mu}\hat{m}_{\nu},
\end{equation}
\begin{align}
    \hat{k} &= -\hat\Omega =  -(\sin\theta\cos\varphi)\hat{x} -
    (\sin\theta\sin\varphi)\hat{y}-(\cos\theta)\hat{z}, \nonumber\\
    \hat{m} &= -(\sin\varphi)\hat{x} + (\cos\varphi)\hat{y}, \nonumber\\
    \hat{n} &= -(\cos\theta\cos\varphi)\hat{x} - (\cos\theta\sin\varphi)\hat{y}-(\sin\theta)\hat{z}, 
    \label{eq:polvectors}
\end{align}
where $\hat\Omega$ is the GW origin direction, and $\hat{x}$, $\hat{y}$, $\hat{z}$ are the Cartesian unit vectors. The equatorial declination and right ascension $(\delta,\alpha)$ are related to the spherical polar coordinates by $\theta=\pi/2-\delta$ and $\phi = \alpha$, where the north celestial pole is in the $\hat{z}$ direction, and the vernal equinox is in the $\hat{x}$ direction. From this point onwards we consider only $\hat\Omega$ and the origin location of GWs.

For a pulsar in direction $\hat{p}$, the antenna response function is the contraction of the polarization basis tensor with the impulse response function of an Earth-pulsar system to a GW, such that:
\begin{equation}
    \mathcal{F}^A(\hat{p},\hat\Omega)\equiv \frac{1}{2}\frac{\hat{p}^\mu\hat{p}^\nu}{1-\hat\Omega\cdot\hat{p}}e^A_{\mu\nu}(\hat\Omega),
\end{equation}
which for each polarization is:  
\begin{align}
    \mathcal{F}^{+}(\hat{p}, \hat{\Omega}) &= \frac{1}{2}
    \frac{(\hat{m}\cdot\hat{p})^{2}-(\hat{n}\cdot\hat{p})^{2}}{1-\hat{\Omega}\cdot\hat{p}}, \nonumber\\
    \mathcal{F}^{\times}(\hat{p}, \hat{\Omega}) &=
    \frac{(\hat{m}\cdot\hat{p})(\hat{n}\cdot\hat{p})}{1-\hat{\Omega}\cdot\hat{p}}.
    \label{eq:antennapatterns}
\end{align}

At a given frequency, the stochastic gravitational-wave background can be written as a superposition of plane waves, such that: 
\begin{align}   \label{eq:arbgwb}
    \quad h_{\mu\nu}&(f, \mathbf{x}) = \nonumber\\ 
    \int_{S^2} &\! \mathrm{d}\hat{\Omega}\left\{ h_{+}(\hat{\Omega}, f)e^{+}_{\mu\nu}(\hat{\Omega}) + h_{\times}(\hat{\Omega}, f)e^{\times}_{\mu\nu}(\hat{\Omega}) \right\} \mathrm{e}^{2\pi\mathrm{i} f \hat{\Omega}\cdot\mathbf{x}/c},
\end{align}
where $c$ is the speed of light, and we integrate over the whole sky $S^2$. In pulsar timing, the integrated effect of the GWs on the pulsar times of arrival (TOAs) depends on the difference in the metric at the Earth (for convenience, the Solar System Barycenter (SSB)) and at the position of the pulsar \citep{abc+09,bf10}. 
For a pulsar $p$ in direction $\hat{p}$, the two contributions to the redshifted arrival rate of TOAs are known as the \emph{Earth term}, $s^{E}_{p}$, and the \emph{pulsar term}, $s^{P}_{p}$. These can be written in the frequency domain as (see \citet{grt+14} for full details):
\begin{align} \label{eq:earthpulsarterm}
    s^{E}_{p}(f) &= \sum_{A=+,\times} \int_{S^2} \! \mathrm{d}\hat{\Omega} \,
    \mathcal{F}^{A}(\hat{p},\hat{\Omega}) h_{A}(\hat{\Omega},f), \nonumber\\
    s^{P}_{p}(f) &= -\!\!\!\!\sum_{A=+,\times} \int_{S^2} \! \mathrm{d}\hat{\Omega} \,
    \mathcal{F}^{A}(\hat{p},\hat{\Omega}) h_{A}(\hat{\Omega},f)
    \mathrm{e}^{-2\pi\mathrm{i} f L_{p}(1-\hat{\Omega} \cdot \hat{p})/c}
\end{align}
where $L_{p}$ is the distance to pulsar $p$, and the total signal is $s=s^{E}+s^{P}$. Appealing to stationarity of the GW background signal, the pulsar term is identical to the Earth term at each frequency except for a phase change, such that $h_{A,p}(\hat\Omega,f)=h_{A,e}(\hat\Omega,f)\equiv h_A(\hat\Omega, f)$. Due to the size of $L_{p}$, the pulsar term oscillates very rapidly across the sky, and much faster than our typical angular resolution. For all practical purposes in stochastic background analyses, we therefore regard the pulsar term phase as random.

Although it is possible to work out some of the integrals of \autoref{eq:earthpulsarterm} analytically for various parametrized power representations on the sky \citep{grt+14}, we can achieve complete generality by re-phrasing the signal response as discretized sums over sky pixels using HEALPix\footnote{Hierarchical Equal Area isoLatitude Pixelation; \url{http://healpix.sourceforge.net}}.  
With that, our Earth term GW sky $h_{A}(\hat\Omega)$ now becomes $h_{(aA)}$, where the index $a$ refers to the $a$-th pixel in the sky, at location $\hat{\Omega}_{a}$. The antenna pattern functions $\mathcal{F}^{A}(\hat{p}, \hat{\Omega})$ become $F_{p(aA)}$, where $p$ is the pulsar index as before. Henceforth, we use the following notation for indices: $(a, b)$ refer to HEALPix pixel indices; $(A,B)$ refer to GW polarization; and $(i,j)$ refer to combined pixel-polarizations, where $i=(aA)$. The matrix $F=\{F_{pi}\}$ is what we refer to as the signal response matrix, and the vector $\mathbf{h} = \{h_{i}\}$ is the full GW sky\footnote{Although we have $2N_\mathrm{pix}$ degrees of freedom in $\mathbf{h}$, \citet{grt+14} showed that half of the GW sky is invisible to pulsar timing arrays, as all the curl modes have vanishing response.}. As for the GW signal, we split up the signal response matrix into an Earth term $F^{E}$ and a pulsar term $F^{P}$. Since the extra phase of the pulsar term is a constant, we absorb it in the signal response matrix such that:
\begin{align} \label{eq:signalresponseEP}
        F^{E}_{p(aA)} &= \sqrt{\frac{3}{2}}
        \frac{\mathcal{F}^{A}(\hat{p},\hat{\Omega}_{a})}{N_\mathrm{pix}} \\
        F^{P}_{p(aA)} &= F^{E}_{p(aA)}
        \mathrm{e}^{-2\pi\mathrm{i} f L_{p}(1-\hat{\Omega}_{a} \cdot \hat{p})/c},
\end{align}
where $N_{\mathrm{pix}}$ is the total number of sky pixels, and we have included a factor of $\sqrt{3/2}$ for a normalization convention that will become apparent in Sec.~\ref{sec:hdcurve}. We can thus write Eq.~\eqref{eq:earthpulsarterm} in matrix notation:
\begin{equation}
    \mathbf{s}(f) = \mathbf{F}\mathbf{h}(f),
    \label{eq:signalresponsepsrpix}
\end{equation}
with $\mathbf{F} = \mathbf{F}^{E} + \mathbf{F}^{P}$ as before. In this single frequency formalism, the size of $\mathbf{s}$ is the number of pulsars $N_\mathrm{psr}$, the size of $\mathbf{h}$ is \textit{twice} the number of pixels (to account for the two polarizations), and thus the size of $\mathbf{F}$ is $(N_\mathrm{psr}\times~2N_\mathrm{pix})$. The array response matrix $\mathbf{F}$ is a constant design matrix for all types of GWB anisotropy. By choosing a sufficient HEALPix resolution, the results from this discretization are indistinguishable from any analytic results. Unless otherwise explicitly stated, in the following we set the HEALPix $N_\mathrm{side}=32$, which gives $N_\mathrm{pix}=12288$. This is fine enough to provide us with fast and accurate numerical solutions.

\subsection{Overlap reduction functions} \label{sec:orf}

For a wide-sense stationary Gaussian GW background, we can write
\begin{equation}
    \langle h_{(aA)} h^{\dagger}_{(bB)} \rangle = \delta_{ab}\delta_{AB} \tilde{P}_{(aA)} = \delta_{ab}\delta_{AB}\frac{P_a}{2},
    \label{eq:correlationgwb}
\end{equation}
where we denote ensemble averaging with $\langle \ldots \rangle$, Hermitian transposition with $\dagger$, and $\delta_{(\ldots)}$ is the Kronecker delta. The GW power at location $\hat{\Omega}_a$ in polarization $A$ is $\tilde{P}_{(aA)}$. We also assume that $\tilde{P}_+ = \tilde{P}_\times = P/2$, where $P$ is the total angular power summed over both polarizations. In all of the following the factor of $1/2$ has already been implicitly absorbed into the pre-factor of \autoref{eq:signalresponseEP}.
    
The overlap reduction function $\Gamma$ of the GW signal is defined as the reduction in signal correlation between two separated detectors \citep{f93}, which in our case corresponds to different Earth-pulsar systems: $\Gamma_{op} = \langle s_p s_o^\dagger \rangle$. Using the signal response matrix formalism of \autoref{eq:signalresponsepsrpix}, the ORF is trivially constructed from \autoref{eq:signalresponsepsrpix} and \autoref{eq:correlationgwb}:
\begin{equation}
    \Gamma_{op} = \sum_{i}  F_{oi} P_i F_{pi}^{\dagger}.
    \label{eq:orf}
\end{equation}
Note that the Earth term and pulsar term components are numerically orthogonal to each other due to the rapid oscillation of the pulsar term across the sky with respect to the Earth term. For the same reason, the pulsar-term contribution to the overlap reduction function is effectively diagonal, meaning we can simply double the diagonal Earth-term ORF components to successfully replicate the full ORF. As such, we have:
\begin{align}  \label{eq:orfep}
    F_{oi}P_{i}F^{\dagger}_{pi} &= F^{E}_{oi}P_{i}F^{E\dagger}_{pi}+\delta_{oi}F^{P}_{oi}P_{i}F^{P\dagger}_{pi} \nonumber\\
    &= (1+\delta_{op}) \times F^{E}_{oi}P_{i}F^{E\dagger}_{pi},
\end{align}
which we can write in compact matrix form as:
\begin{align} \label{eq:orf_matprod}
    \mathbf{\Gamma}^E &= \mathbf{F}^E \cdot \mathbf{P} \cdot \mathbf{F}^{E,\mathrm{T}}, \nonumber\\
    \Gamma_{pp} &= 2\Gamma^E_{pp}, \quad \Gamma_{op} = \Gamma^E_{op},
\end{align}
where $\mathbf{\Gamma}$, $\mathbf{\Gamma}^E$ is the $(N_\mathrm{psr}\times~N_\mathrm{psr})$ array ORF matrix for the full signal and Earth-term-only signal, respectively. We also note that normal transposition has replaced Hermitian transposition since $\mathbf{F}^E$ is real.  
The diagonal $(2N_\mathrm{pix}\times~2N_\mathrm{pix})$ matrix $\mathbf{P}$ contains the array of power values in each sky pixel along the diagonal, with $+,\times$ polarizations as pixel neighbors.  

\subsection{The Hellings \& Downs curve} \label{sec:hdcurve}

As a sanity check, we recover the ORF for an isotropic background (known as the \textit{Hellings and Downs curve} \citep{hd83}), by modeling $\mathbf{P} = \mathbb{I}_{2N_\mathrm{pix}}$.  
For an arbitrary array of pulsars we can construct the signal response matrix and carry out the matrix operations in \autoref{eq:orf_matprod}, which correspond to a double integral over the sky. When integrating analytically, the resulting ORF expression is a function only of the angular separation of pulsars rather than their absolute position.  
However, tackling this integral analytically requires a combination of coordinate transformations and contour integration \citep{abc+09,msmv13,grt+14}, all of which has been avoided by casting the problem into matrix algebra.
    
\begin{figure} 
\centering
\includegraphics[width=0.5\textwidth]{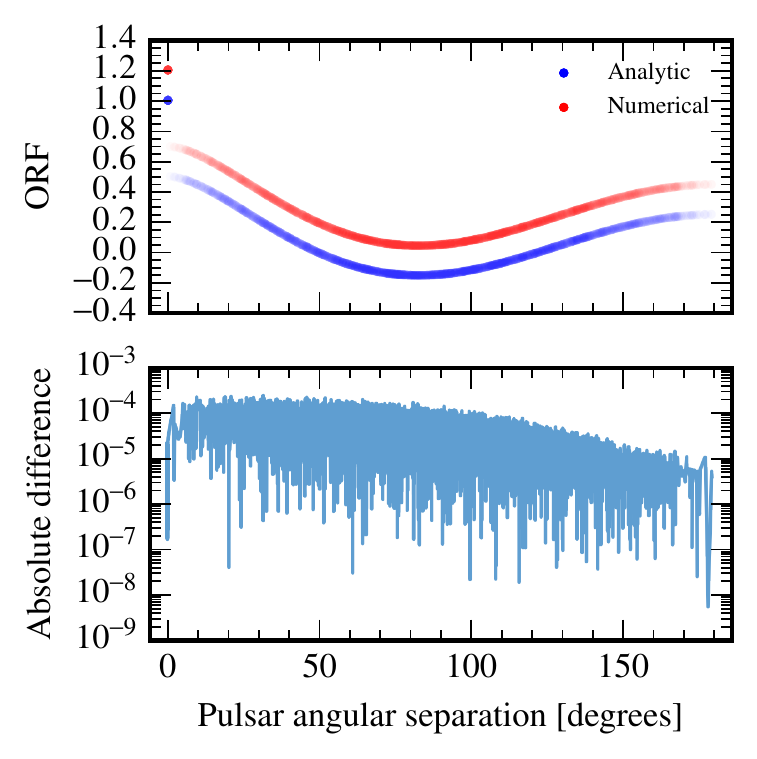}
\caption{Comparison of analytic and numerical \textit{Hellings and Downs curve} computation for a $100$-pulsar array. The analytic calculation uses \autoref{eq:hdcurve}, while the numerical calculation assumes an isotropic power-map within \autoref{eq:orf_matprod}. In the top panel, the numerical calculation is offset from the analytic one by $+0.2$ for ease of viewing. The lower panel shows the absolute difference between the numerical and analytic calculations.}
 \label{fig:anis_numaccuracy}
\end{figure}

For an array of $100$ pulsars drawn from an isotropic distribution on the sky, we compare our numerical results to the following analytic expression (which is normalized by convention to equal $1$ for $o=p$, thus explaining the $\sqrt{3/2}$ factor adopted in \autoref{eq:signalresponseEP}):
\begin{equation}
    \Gamma(\xi_{op}) = \frac{3}{2}x\ln(x)-\frac{x}{4}+\frac{1}{2}+\frac{\delta_{op}}{2}, \quad x = \frac{1-\cos\xi_{op}}{2},
    \label{eq:hdcurve}
\end{equation}
where $\delta_{op}$ accounts for the pulsar term contribution. This comparison is shown in \autoref{fig:anis_numaccuracy}, where in the top panel the numerical ORF values are offset from the analytic ORF values by $+0.2$ for ease of viewing. The lower panel shows that the absolute difference between the numerical and analytic values remains $\lesssim 10^{-4}$ over the entire angular separation range. Our numerical ORF formalism is fast, flexible, and accurate. 
    
\subsection{GW map parametrizations} \label{sec:map_models}

In a pixelated power representation such as ours, the most obvious parametrized power representation is to search for the $P_i$ values individually at a pre-determined pixel resolution. This is an inelegant and inefficient approach though, since the number of parameters in the model gets prohibitively large, and PTAs typically do not have a high response resolution. In the following we list several parametrizations that are appropriate for different signal scenarios. One of these has appeared already in the literature (the spherical-harmonic model) while the rest are new to this paper. While we discuss these parametrizations separately, the models can be linearly combined where necessary to fit broad anisotropic features (spherical-harmonic or disk) and localized features (point sources).

\subsubsection{Spherical harmonic modeling of power}

The spherical-harmonic basis is an obvious choice to decompose a scalar-field on the $2$-sphere, since the lowest order basis function describes isotropy, and higher functions add anisotropic power at smaller angular scales. 
Spherical-harmonic power reconstructions were first proposed in this context for LIGO sky mapping \citep{ao97}, but have recently received significant attention in the PTA literature \citep{msmv13,tg13,grt+14,tmg+15}, where the power is decomposed as:
\begin{equation}
    P(\hat{\Omega}) = \sum_{l=0}^{\infty}\sum_{m=-l}^{l} c_{lm}
    Y_{lm}(\hat{\Omega}),
    \label{eq:sphdec}
\end{equation}
where $Y_{lm}$ are the real-valued spherical harmonics, and $c_{lm}$ are the spherical harmonic coefficients of $P(\hat{\Omega})$. In this parametrization, an isotropic sky is described by $\{c_{00}=\sqrt{4\pi}\}$, while a maximal dipole in the $+\hat{z}$ direction is described by $\{c_{00}=\sqrt{4\pi}, c_{10}=\sqrt{4\pi/3}\}$.
This linear parameterization of the power in terms of the $c_{lm}$ coefficients leads to a similar expression for the ORF:
\begin{equation}
    \Gamma_{op} = \sum_{l=0}^{\infty}\sum_{m=-l}^{l} c_{lm} \Gamma_{(lm)(op)},
    \label{eq:aniorf}
\end{equation}
where $\Gamma_{(lm)(op)}$ are the anisotropy ORF components for anisotropy components $(l,m)$ between pulsars $p$ and $o$, which can be pre-computed for a specific PTA from just the pulsar sky locations.
    
Although \citet{msmv13} and \citet{grt+14} have given detailed analytic formulae to evaluate the $\Gamma_{(lm)(op)}$ for any $(lm)$, in practice it is equally efficient to use linear algebra to compute these quantities numerically using \autoref{eq:orf}:
\begin{equation}
    \Gamma_{(lm)(op)} = \sum_{i} c_{lm} Y_{(lm)i} F_{oi} F_{pi}^{\dagger},
    \label{eq:numaniorf}
\end{equation}
where $Y_{(lm)i}$ is the value of $Y_{(lm)}$ evaluated at sky location $\hat{\Omega}_i$. In this formalism, the model parameters are the coefficients $c_{lm}$. Normalization is ensured by fixing $c_{00}=\sqrt{4\pi}$ since the $l\geq 1$ spherical-harmonics are orthonormal to the constant $l=0$ function (where $Y_{00}=1/\sqrt{4\pi}$), and thus themselves integrate to zero across the sphere. Since $c_{00}$ is fixed, variation of the monopolar level of the GWB is absorbed into the strain amplitude parameter.
    
For a given set of $\{c_{lm}\}$ to represent a physical power map, the resulting power has to be non-negative at every point on the sky. In practice this is performed by evaluating \autoref{eq:sphdec} on a grid of test sky-positions for a set of proposed $\{c_{lm}\}$ \citep{tg13}. In this paper we sample uniformly in $c_{lm}\in U[-5,5]$ where $l\geq 1$, and evaluate the \textit{physical prior} on a HEALPix $N_\mathrm{side}=8$ sky, which gives $768$ test positions. In a model with $l_\mathrm{max}$ as the multipole cutoff, the number of anisotropy parameters is $(l_\mathrm{max}+1)^2-1$. In Section \ref{sec:simulations} we study how sky reconstruction and the Bayes factor for anisotropy varies with the choice of $l_\mathrm{max}$, up to $l_\mathrm{max}=6$.
    
\subsubsection{Spherical-harmonic modeling of square-root of power} \label{sec:sqroot_spharm}

While the spherical harmonics are a natural basis with which to decompose the GWB power, the practical implementation has required us to introduce a non-analytic prior on the $c_{lm}$ coefficients.  
We can bypass this rejection prior by instead paramatrizing the \textit{square root of power}, which is allowed to be negative, and will by construction always produce a positive power distribution. The benefits are twofold -- $(i)$ the prior is now completely analytic; and $(ii)$ sampling performance is improved through better mixing, shorter burn-in, and the potential to propose draws from the prior volume during sampling\footnote{The ability to implement draws from the prior aided particularly in sampling posterior multi-modalities that derived from a given $P(\hat\Omega)$ having two possible differing-sign solutions for $P(\hat\Omega)^{1/2}$.}. 

We decompose the square root of power in exactly the same way as for the power:
\begin{equation}
    P(\hat{\Omega})^{1/2} = \sum_{L=0}^{\infty}\sum_{M=-L}^{L} a_{LM} Y_{LM}(\hat{\Omega}),
    \label{eq:}
\end{equation}
where $Y_{LM}$ are the real-valued spherical harmonics, and $a_{LM}$ are search coefficients. We have capitalized the component labels $\{LM\}$ to be distinct from the usual $\{lm\}$ components of the GWB power. 
    
There is not a direct mapping between structure in $l$ of the power and structure in $L$ of the root-power\footnote{Although both low-$L$ and low-$l$ behaviour describe broad anisotropic features.}, e.g. a dipole in power will be described by $l_\mathrm{max}=1$, but in principle requires contributions in the root-power from $L>1$. We show this analytically in Appendix \ref{sec:appA}, where the power coefficient $c_{lm}$ is decomposed as an infinite summation of $a_{LM}$ terms. However, we also show that adequate reconstruction of a map with $l\leq l_\mathrm{max}$, can be achieved with $L\leq L_\mathrm{max}=l_\mathrm{max}$.\footnote{After submitting this article, a similar method has been proposed in the context of ground-based GW constraints on the anisotropy of BBH merger events \citep{2020arXiv200611957P}.} 
    
\subsubsection{Point-source anisotropy} \label{sec:pntsrc}

The spherical-harmonic model builds towards smaller-scale anisotropic structure through successively higher multipoles. Thus if the GW sky is predominantly isotropic but with one bright SMBBH sticking prominently above the background level, we will need a high $l_\mathrm{max}$ (and consequently a large number of parameters) to localize it. Instead, we can model an isolated point of GWB power with a 2-sphere delta-function, such that $P(\hat\Omega) = P'\delta^2(\hat\Omega,\hat\Omega')$. The corresponding physical picture is of a \textit{stochastic point source} \citep{2006CQGra..23S.179B,2008PhRvD..77d2002M,cs13}. 

We implement this point source of power by lighting up a single pixel in our modeled sky, and keeping all other pixels dark. The only search parameters are for the sky location of the lit pixel, which can be referred to through pixel indices. 
The power in this pixel is set to $N_\mathrm{pix}$ to satisfy normalization, and the ORF is computed using \autoref{eq:orf_matprod}. The source sky location is searched over with a uniform prior on the sky, i.e. $\alpha\in U[0,2\pi]$, $\cos(\pi/2-\delta)\in U[-1,1]$. We can combine these point sources with different amplitude weightings (sometimes in addition to an isotropic background) to model whatever small-scale angular features are supported by the data. 

In Section \ref{sec:simulations} we study how sky reconstruction and the Bayes factor for anisotropy varies as we model the sky with different numbers of point sources, from $N_\mathrm{point}=1,\ldots,5$. The scenario of multiple modeled point sources stacking on top of each other during sampling is avoided through the natural parsimony of Bayesian model selection. We help this along during the posterior sampling by implementing a prior that tests and rejects any new proposed point-source location that would be cast to the same pixel index as another in an $N_\mathrm{side}=8$ sky. 
We note that combining this point-source model of anisotropy with a frequency-dependent amplitude parameter permits a power-based search for bright single GW sources that can complement existing continuous-wave and burst methods.
    
\subsubsection{Disk anisotropy} \label{sec:diskanis}

If the angular power distribution reflects the local clustering of host galaxies to produce an extended \textit{hotspot} region on the sky, then both the spherical-harmonic and point-source models have drawbacks. We can combine the ability of the spherical-harmonic approach to model broad anisotropic features with the ability of the point-source model to localize bright regions of power. This hybrid is a \textit{disk} of GW power with a modeled sky location and angular radius. The power is constant inside the disk perimeter and zero outside. This is implemented using the HEALPix \texttt{query\_disc} function, which returns an array of pixel indices lying within a disk of supplied sky location and angular radius.  
The disk model has three search parameters: two for sky-location (with a uniform prior on the sky) and one for the angular radius (with prior $\theta_r\in U[0,\pi]$ to avoid the disk wrapping around the sphere and overlapping with itself).

\section{Likelihood}
\label{sec:likelihood}

We significantly simplify the pulsar-timing likelihood to permit fast analysis of datasets with a wide range of different mapping parametrizations. We are only interested in the comparative modeling and detection efficacy of different GWB map parametrizations, and so ignore the pulsar ephemeris model. The dominant effect of this ephemeris model on signal inference is to remove sensitivity to lower frequencies due to the subtraction of a quadratic function (which models the pulsar spin-down). It also diminishes sensitivity at sampling frequencies of $1\,\mathrm{yr}^{-1}$ and $2\,\mathrm{yr}^{-1}$ due to the fitting of astrometric parameters. Likewise, sophisticated noise modeling is essential to properly isolate GW signals from noise processes, but we do not include processes beyond radiometer noise here. Nevertheless, we stay true to the spirit of the
likelihood as it is used in real searches (e.g., \citep{abb+15,abb+16}),
while stripping out all the model contributions that are irrelevant to this paper. 

\subsection{Assumptions} \label{sec:like_assumptions}

In our simplified model, we have $N_p$ pulsars with $N_\mathrm{obs}$ observations each.  For pulsar $p$ the observations $\delta\mathbf{t}_{p}$ consist of two components:
 \begin{equation}
    \delta\mathbf{t}_{p} = \mathbf{s}_{p} + \mathbf{n}_{p},
    \label{eq:residuals}
\end{equation}
where $\mathbf{s}_{p}$ is the GW signal and $\mathbf{n}_{p}$ is the noise. Given our focus on GWB map parameterizations, we assume a simple Gaussian white-noise model in our simulations: $\langle n_{pc} n_{od} \rangle = \sigma_{p,c}^2 \delta_{op} \delta_{cd} = N_{p,cd}\delta_{op}\delta_{cd}$, where $\{p,o\}$ index pulsars, $\{c,d\}$ index observations, and $\sigma_{p,c}$ is the uncertainty in a single observation. This uncertainty is usually referred to as the radiometer noise \citep{abb+15}, which we here assume to be a known homoscedastic quantity for each pulsar. The $\langle \ldots \rangle$ indicates ensemble averaging (equal to the time average here since we assume ergodicity).

We model the GW signal, $\mathbf{s}_{a}$, as a zero-mean, wide-sense stationary stochastic background 
with the covariance between observation $c$ in pulsar $p$ and observation $d$ in pulsar $o$ parametrized as:
\begin{equation} \label{eq:gwprior}
    \langle s_{pc} s_{od} \rangle = \Gamma_{po}C(|t_c-t_d|) ,
\end{equation}
where $\Gamma_{po}$ is the ORF between these two pulsars (i.e. the spatial correlation), and $C(|t_c-t_d|)$ is the time-domain correlation induced by the GWB. 
    
For the sake of simplicity we further assume that the GWB signal has been whitened, such that $C(|t_c-t_d|)=A^2\delta_{cd}$. This assumption in no way affects the interpretation of our results in this paper, since we are only interested in the relative efficacy of power-map parametrizations, and not spectral parametrizations. The true GWB from a population of SMBHB will induce long-timescale stochastic deviations in the TOAs, with an $f^{-13/3}$ power-law spectral density \citep{rm95,wl03,jb03,bbr80,p01} that corresponds to a dense time-domain covariance matrix. This means that in real searches most information about the GWB is derived from the lowest few sampling frequencies in the TOA data. In such real analyses the dense time-domain covariance is modeled through rank-reduced means \citep{lha+13,vhv14,vhv15}, or through data whitening based on estimates of the covariance \citep{chc+10}. Assuming a white GWB spectrum here allows us to rapidly compute the PTA likelihood. However, our modeling of anisotropy and the PTA overlap reduction function requires very little effort to incorporate into real production-level pipelines that deal with multiple GW frequencies and unevenly-sampled time-series data, which we discuss in Sec.~\ref{sec:conclusions}. 

\subsection{Marginalized likelihood}
    
We marginalize over the low-level linear parameters $\mathbf{s}_{p}$ and $\mathbf{n}_p$ with Gaussian priors \citep{lha+13,abb+15,abb+16,vhv14}. This gives a likelihood that depends only on model hyperparameters: the GWB amplitude $A$; the TOA uncertainties $\{\sigma_p\}$ (assumed to be the same for all observations and all pulsars, with a delta-function prior set at the measured uncertainties); and the GWB map parameters $\{c_\mathbf{P}\}$. The log-likelihood $\ln\mathcal{L}=\ln p(\delta\mathbf{t}|A,\sigma_p,c_\mathbf{P})$ for all pulsars is then given by:
 \begin{equation} \label{eq:loglik}
    \ln\mathcal{L} = \sum_{c=1}^{N_\mathrm{obs}}\bigg\{- \frac{1}{2}\delta\mathbf{t}_{c}^{T}\mathbf{\Sigma}_c^{-1}\delta\mathbf{t}_{c} - \frac{1}{2}\ln\det 2\pi\mathbf{\Sigma}_c \bigg\},    
\end{equation}
where $\delta\mathbf{t}_{c}$ is a vector of size $N_p$ consisting of the $c$-th observation for each pulsar, and
\begin{equation}
    \mathbf{\Sigma}_c = A^2\mathbf{\Gamma} + \mathbf{N}_c, 
    \label{eq:sigma}
\end{equation}
where $\mathbf{\Sigma}_c$ is the $(N_p\times~N_p)$ inter-pulsar correlation matrix of the $c$-th observation, $\mathbf{\Gamma}$ is the $(N_p\times~N_p)$ dense ORF matrix of the $c$-th observation (we assume that it is the same for all observations), and $\mathbf{N}_c$ is the $(N_p\times~N_p)$ diagonal noise covariance matrix of the $c$-th observation. The latter noise matrix is assumed to be the same for all pulsars and all observations, and remains fixed in our sampling. In production-level PTA analysis pipelines, the covariance matrices of long-timescale stochastic processes (like the GWB) are modeled on rank-reduced bases \citep{vhv14,vhv15,lha+13}, e.g. a Fourier basis. It is thus straightforward to generalize our ORF modeling to permit different ORFs at each frequency. Additionally, generalizing \autoref{eq:loglik} to account for data taken at different times in each pulsar is most easily achieved by concatenating all observations into a single data vector, with $\mathbf{N}$ still diagonal over the entire PTA, and $A^2\mathbf{\Gamma}$ now equal to the full-array version of \autoref{eq:gwprior}. 
   
We apply a uniform prior in $\log_{10}(A\, /\, \mathrm{secs})$ over a sampling range $[-10,-3]$. All remaining variables in the model are used to parametrize the GWB map, except in the case of multiple point sources, where each component, $k$, receives a separate amplitude parameter $A_k$ to permit different source weightings. The priors on $\{c_\mathbf{P}\}$ are listed in Sec.\ \ref{sec:anisotropy} for each parametrization.  
The $A^2\mathbf{\Gamma}$ term describes the temporal and spatial correlations induced by the GWB, which can be expressed through linear combinations of anisotropic features using the parametrized models in this paper, e.g. an isotropic background with multiple bright binaries has $A^2\mathbf{\Gamma} = A_\mathrm{iso}^2\mathbf{\Gamma}_\mathrm{iso} + \sum_{k=1}^{N_\mathrm{comp}}A_k^2\mathbf{\Gamma}_k$.

\section{Simulations}
\label{sec:simulations}

\subsection{Tested GW sky maps} \label{sec:setup}

We test the relative efficacy of our mapping parametrizations on two qualitatively different types of GW skies, shown in \autoref{fig:test_gwskies} with their associated angular power spectrum. These are the following:
\begin{figure*}[htb]
\subfloat[\label{subfig:a}]{%
  \includegraphics[width=0.325\textwidth]{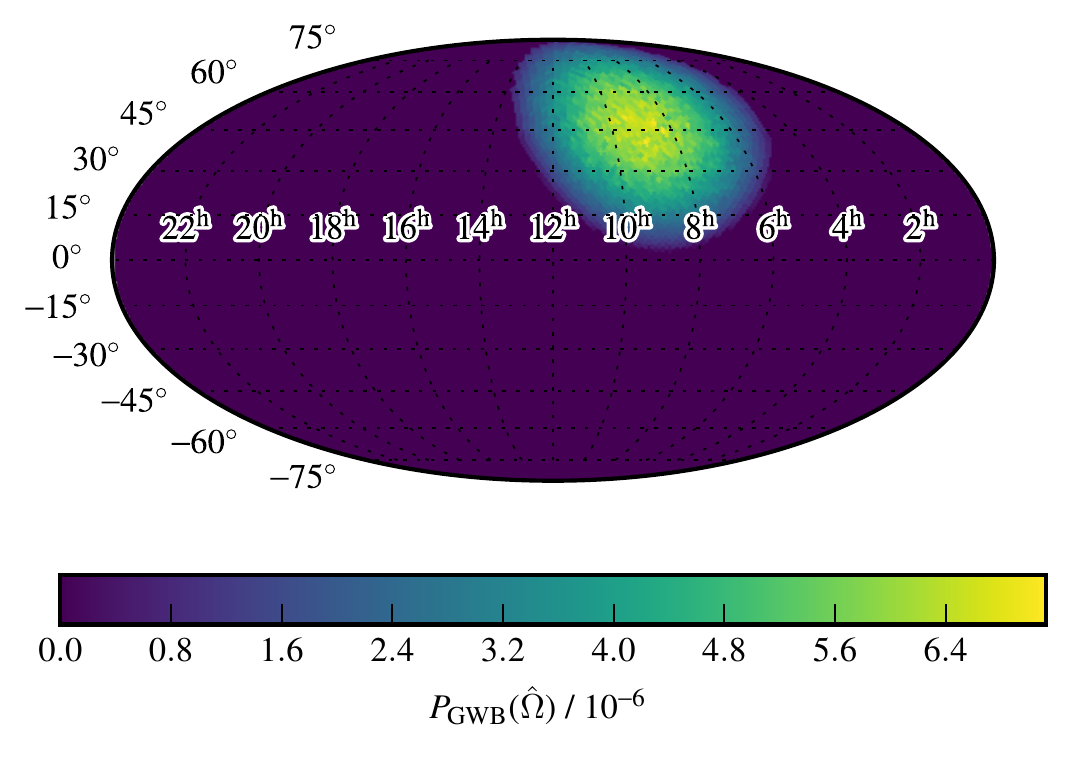}%
}
\subfloat[\label{subfig:b}]{%
  \includegraphics[width=0.325\textwidth]{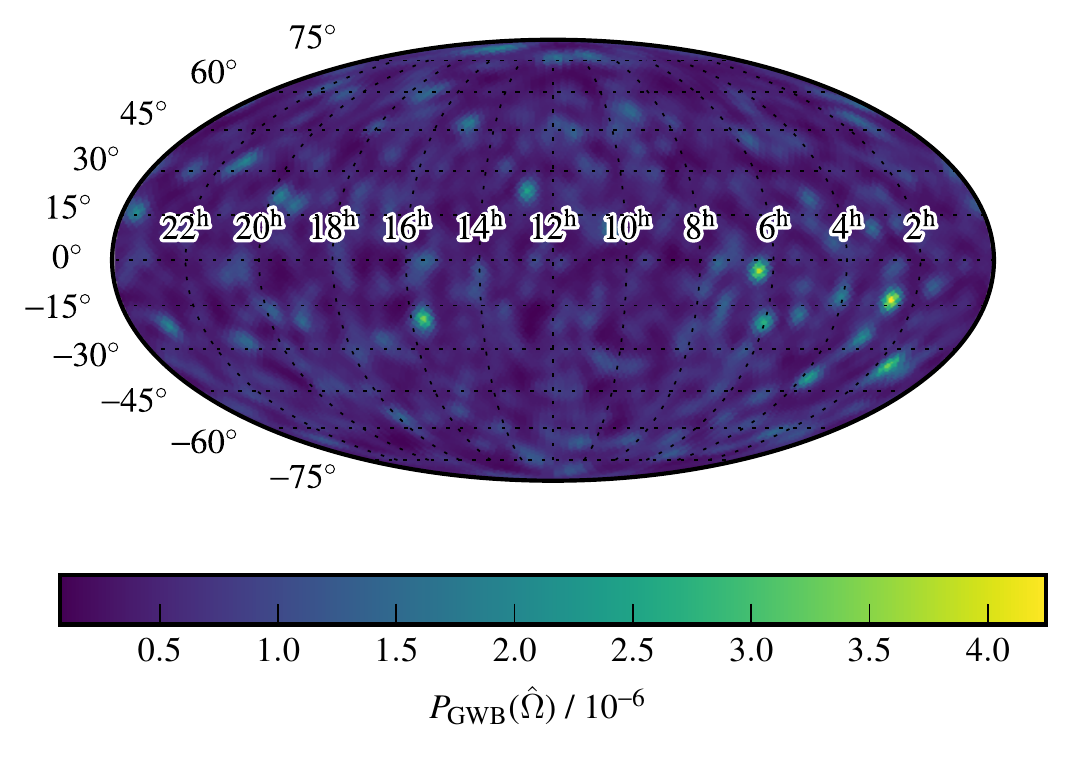}%
}\hfill
\subfloat[\label{subfig:c}]{%
  \includegraphics[width=0.325\textwidth]{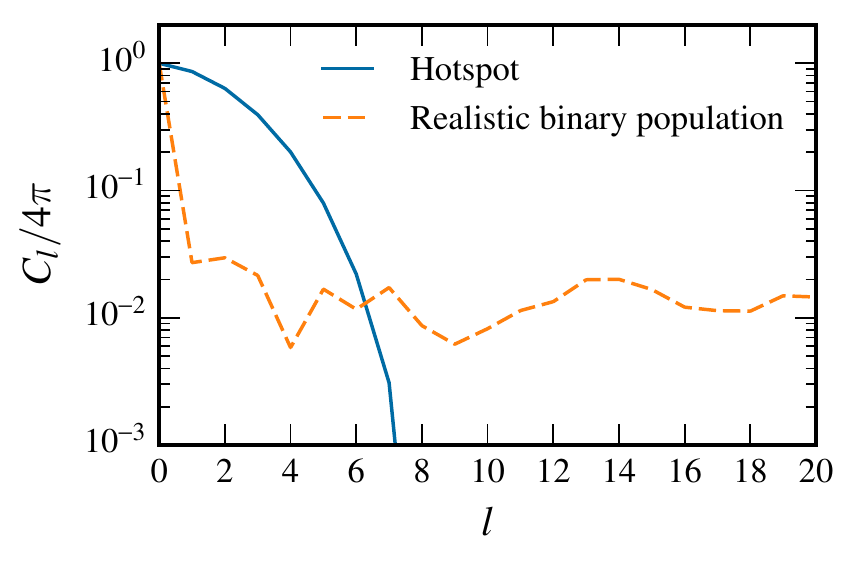}%
}
\caption{The GW skies that we test our models against in this paper. The map in (a) represents a ``hotspot'' region of excess GW power, while the map in (b) is a smoothed  map generated from a synthesized population of SMBBHs. The angular power spectrum of these maps is shown in (c).}
\label{fig:test_gwskies}
\end{figure*}
\begin{description}
    \item [Hotspot] an extended hotspot of power that dominates the entire GW sky. This is an edge-case that in less extreme scenarios may represent a cluster of galaxies hosting many merging SMBBHs, or a local Universe inhomegenity in the SMBBH distribution. This map was constructed by performing an $l\leq5$ decomposition on an isotropic map with a large value added to one pixel. The extended region of power was excised and added to an otherwise empty map.
    \item [SMBBH population] a more realistically diffuse GW sky created from the output of a SMBBH population synthesis simulation constructed with the technique described in \citep{s13}. The SMBBH population is constructed empirically based on a combination of observed quantities and theoretical input. In short, a galaxy merger rate is constructed by multiplying the galaxy mass function to the observed galaxy pair fraction and by dividing it for the characteristic merger timescale, which is estimated from numerical simulations. Galaxies are then populated with SMBHs according to scaling relations drawn from the literature. The model adopted here combined the galaxy mass function from \cite{2013ApJ...777...18M}, pair fraction from \cite{2009ApJ...697.1369B}, merger timescale from \cite{2010MNRAS.404..575L}, and SMBH-bulge mass relation from \cite{2004ApJ...604L..89H}, to produce a catalog of SMBBHs with chirp masses, redshifts, and orbital frequencies from which the GW strain of each (assumed circular) binary was computed. By selecting those systems with observed GW frequencies between $0.1$~yr$^{-1}$ and $0.2$~yr$^{-1}$, only binaries within the lowest frequency resolution bin of a PTA with $10$-year timing baseline were collected to create the sky map. The angular positions were assigned to be statistically isotropic, but the strain distribution is dominated by only a small number of GW-bright systems, producing anisotropy in the angular power distribution. Most SMBBH systems in this map constitute a background of sources that are not individually resolvable, with only a handful of systems emitting signals that resound prominently above this. The displayed map has been smoothed with a Gaussian kernel to allow structure to be seen; the true map looks mostly isotropic with only a few dominant pixels due to bright systems.
\end{description}

In the following we describe the format of the tested datasets that contain the imprint of these GW power distributions, and the suite of signal injections.

\subsection{Simulation setup} \label{sec:setup}

Our interest lies in the most apt model for GWB anisotropy rather than finding the optimal geometrical properties of a PTA. Hence, we create a statistically isotropic array of $100$ pulsars on the sky, as may be operational in the mid- to late-$2020$s when probes of signal anisotropy will be possible. Current arrays use $\sim 40-50$ pulsars in flagship GW searches \citep{ng12p5,ng12p5_wb} (although NANOGrav now times as many as $79$ pulsars \citep{ngwp19}), so forecasting an additional $\sim 6$ pulsars per year to be added to the IPTA is reasonable \citep{taylor+16}. As discussed in Sec.~\ref{sec:like_assumptions}, we do not concern ourselves with the pulsar timing model, nor differing noise characteristics. The pulsars only differ through their geometrical response to a GW signal, as characterized by their position on the sky. Each pulsar has $100$ observations that are evenly sampled and concurrent across the array, with homoscedastic white-Gaussian noise of standard deviation $100$ ns.  

For reasons discussed in Sec.~\ref{sec:like_assumptions}, the GWB is modeled as a whitened Gaussian process of standard deviation $A$ (which is varied in the injections) that is spatially correlated across the array. This spatial correlation depends on the ORF and thus the angular power distribution of the GW sky. Since the GWB is whitened, a single PTA dataset with $100$ observations per pulsar can be considered as a concatenation of (for example) $20$ separate realizations of a signal with $5$ degrees of freedom (i.e. informative frequencies), which is representative of signals that are likely to be recovered in the near future. Hence we create many datasets with different values of $A$ for each sky, but the random seed used to produce the signal realizations is kept fixed, as is the seed for the noise injection, which is fixed at an independent value for each pulsar. The hotspot datasets have signals injected on an evenly-spaced grid of $50$ values of $A$ in the range $[5,50]$ ns. The realistic SMBBH population datasets have signals injected on an evenly-spaced grid of $100$ values of $A$ in the range $[10,500]$ ns.

\section{Results}
\label{sec:results}

An isotropic background is a convenient baseline model with which we compare the efficacy of different skymap parameterizations. Indeed, for our chosen PTA configuration we can compute the normalized geometrical match between the ORF of our two different skies with the corresponding isotropic ORF. This represents the theoretical limiting match between the two ORFs in the infinite SNR limit, taking the form \citep{cs16,tlb+17}
\begin{equation}
    M = \sum_{p,o\neq p} \frac{\Gamma^\mathrm{sky}_{po} \Gamma^\mathrm{iso}_{po}}{(\Gamma^\mathrm{sky}_{po} \Gamma^\mathrm{sky}_{po})^{1/2} (\Gamma^\mathrm{iso}_{po} \Gamma^\mathrm{iso}_{po})^{1/2}},
\end{equation}
and has values of $M=0.78$ for the hotspot sky and $M=0.98$ for the SMBBH population sky. The signal correlations seen by our PTA are not so different from those induced by an isotropic GWB, even in the case of the hotspot sky.

The aforementioned statistic only accounts for the geometrical match of signal correlations, but the noise properties of the pulsars and the presence of the GW signal in the arrival time deviations should also be accounted for. In fact, introducing signal and noise weightings takes us to the signal-to-noise ratio (SNR). The general form of the expected value of this SNR is \citep{abc+09,sej+13}
\begin{equation} \label{eq:snr}
    \langle\rho\rangle = \left( 2T\sum_{p,o\neq p} \Gamma^\mathrm{true}_{po}\times\Gamma^\mathrm{template}_{po} \int_{f_l}^{f_h} df \frac{P_g^2(f)}{P_p(f)P_o(f)}\right)^{1/2}, 
\end{equation}
where $T$ is the timing baseline, $P_g(f)$ is the spectrum of GWB-induced timing residuals, and $P_{p,o}(f)$ is the spectrum of noise-induced residuals in each pulsar. If radiometer noise is the only source of intrinsic per-pulsar noise, then $P_p(f) = 2\sigma_p^2\Delta t_p + \Gamma_{pp}P_g(f)$, where $\sigma$ is the RMS uncertainty in the timing residuals and $\Delta t$ is the effective timing cadence. \autoref{eq:snr} accounts for the possibility of a template ORF being different from that of the true data. If the template ORF equals the true ORF, then we call $\langle\rho\rangle$ the \textit{optimal} SNR; otherwise if we adopt an isotropic template this is the \textit{suboptimal} SNR. We will now drop the expectation brackets for ease of readibiltiy. The ratio of the suboptimal SNR to the optimal SNR depends mildly on the strength of the GWB signal; in our simulations, the hotspot case has $(\rho_\mathrm{subopt}/\rho_\mathrm{opt})\sim0.59-0.62$, while for the SMBBH population $(\rho_\mathrm{subopt}/\rho_\mathrm{opt})\sim0.96-0.97$.

\subsection{Hotspot sky tests}
In order to guide our intuition in later cases, we use the hotspot sky for a first round of model selection with several of our sky parametrizations. Since the map was initially constructed from an $l_\mathrm{max}=5$  restricted multipole decomposition of a single-pixel sky, we perform a spherical harmonic reconstruction with $l_\mathrm{max}=5$. We also test a disk and point-source model; in both of these cases, the angular power is zero except inside the pixel region of the disk/point. We do not test the $P(\hat\Omega)^{1/2}$ spherical-harmonic decomposition on this hotspot sky, instead deferring this to the SMMBH sky where its full efficacy can be shown. These hotspot sky tests are not meant to be an exhaustive study of all methods, instead acting as a guide for our intuition and a test of our sky-mapping framework. We perform simultaneous Bayesian parameter estimation and model selection using the product-space sampling approach outlined in Appendix \ref{sec:appB}. The relevant Bayes factor that we track is between a given anisotropy model and the baseline assumption of an isotropic sky. 

\begin{figure}
\centering
\includegraphics[width=0.5\textwidth]{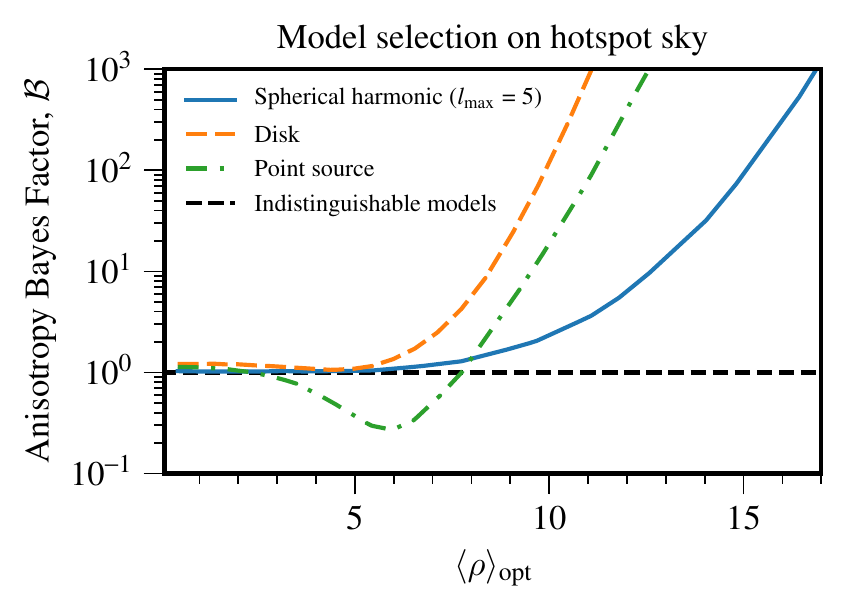}
\caption{The Bayes factor for anisotropy versus isotropy of the hotspot GW sky under different model parametrizations is shown here. The most apt model is clearly the disk and point-source descriptions, although the point source suffers an initial period of being disfavored.}
\label{fig:hotspot_methods}
\end{figure}

\autoref{fig:hotspot_methods} shows that at low SNR the data is initially uninformative and unable to arbitrate any deviation from isotropy, giving $\mathcal{B}\sim1$. As the SNR grows, all models except the point source have monotonically increasing Bayes factors. The initial tendency for the data to favor an isotropic sky over a point source is due to the extreme spatial compactness demanded by this model, with all power in one pixel. The data initially prefer isotropy, but as the SNR grows so does the data's tendency to favor this spatially compact description to explain the hotspot of power. As one would expect for this sky, a disk of power is the favored model overall, striking a balance between appropriateness to the power distribution and model parsimony. The results in \autoref{fig:hotspot_methods} can be compared with those of \citet{hkj19}, where a localized GW beam accounting for more than $50\%$ of the total GW energy density can be detected at $3$-sigma when the SNR$\geq 10$. 

\begin{figure*}[htb]
\subfloat[\label{fig:hotspot_lmax_compare}]{%
  \includegraphics[width=0.492\textwidth]{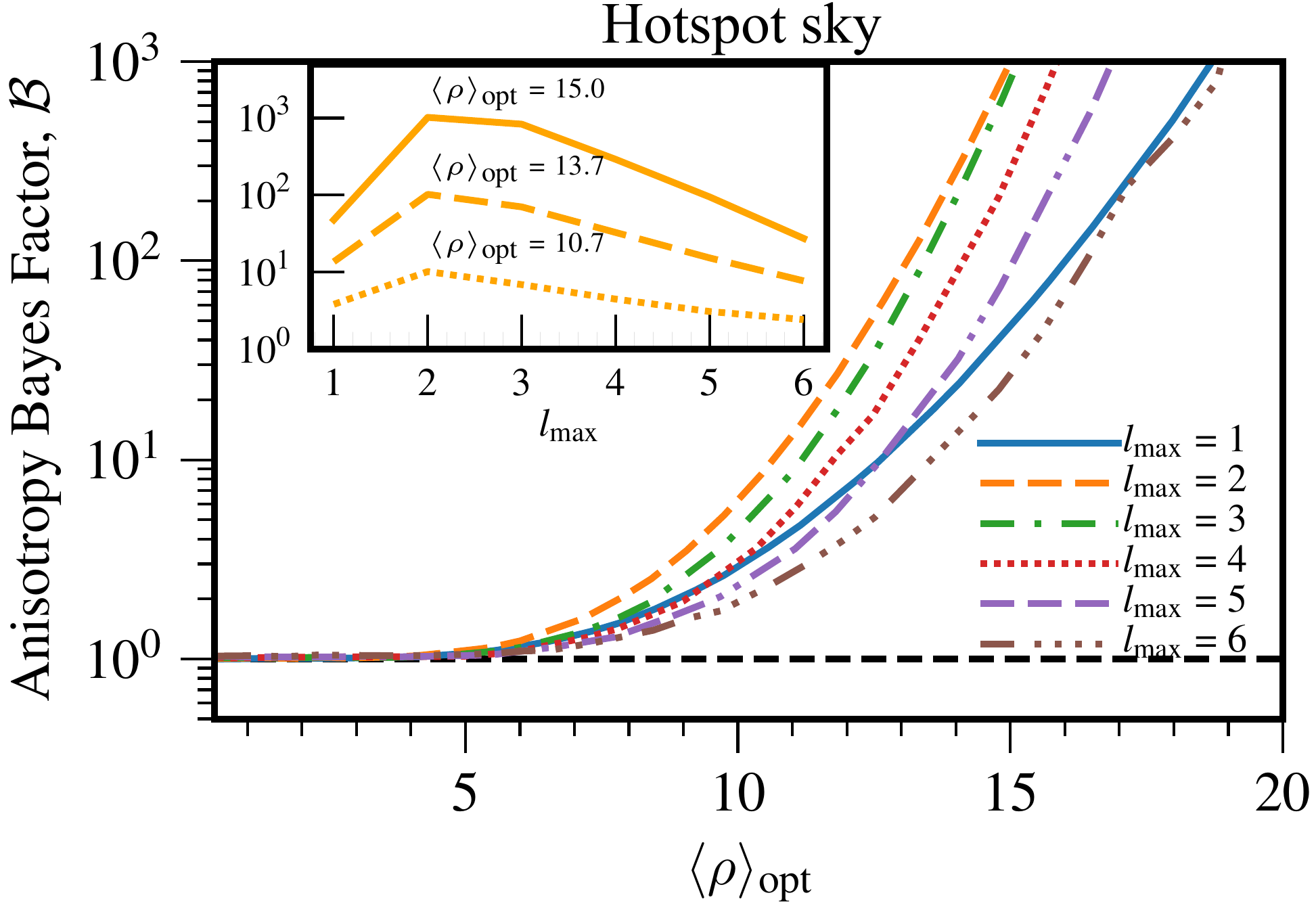}%
}
\subfloat[\label{fig:realpop_lmax_compare}]{%
  \includegraphics[width=0.492\textwidth]{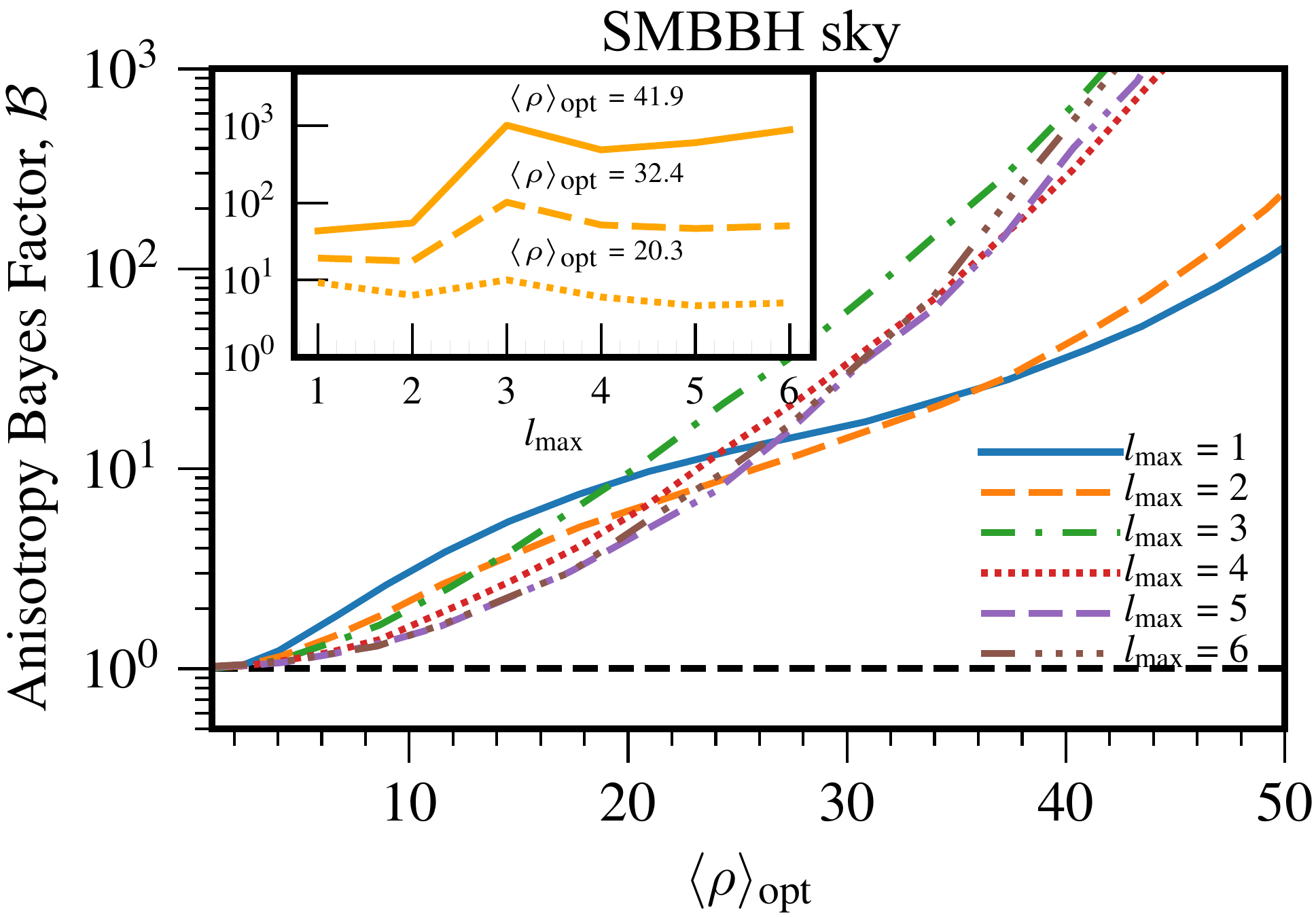}%
}
\caption{Comparison of anisotropy-vs-isotropy Bayes factor growth with SNR for different choices of $l_\mathrm{max}$ in spherical-harmonic power modeling of (a) a hotspot sky, and (b) a realistic SMBBH sky. Inset figures show vertical slices through the curves at different SNR values.}
\label{fig:bfs_spharmonly}
\end{figure*}
While the spherical harmonic model seems to perform the worst overall in \autoref{fig:hotspot_methods}, it has been artificially restricted here to $l_\mathrm{max}=5$. We now relax this, showing in Fig.~\autoref{fig:hotspot_lmax_compare} the Bayes factor growth as a function of SNR and different $l_\mathrm{max}$. The inset figure slices through the different $l_\mathrm{max}$ curves at fixed SNR, showing how the Bayes factor evolves with $l_\mathrm{max}$ under fixed signal strengths. An $l_\mathrm{max}=2$ model is favored at all SNRs, although there is a hint that $l_\mathrm{max}=3$ may be equally favored at higher SNRs, where more informative data counteracts the Occam penalty brought by more model parameters. 

\begin{figure*}
\subfloat[\label{fig:hotspot_recongallery}]{%
  \includegraphics[width=0.492\textwidth]{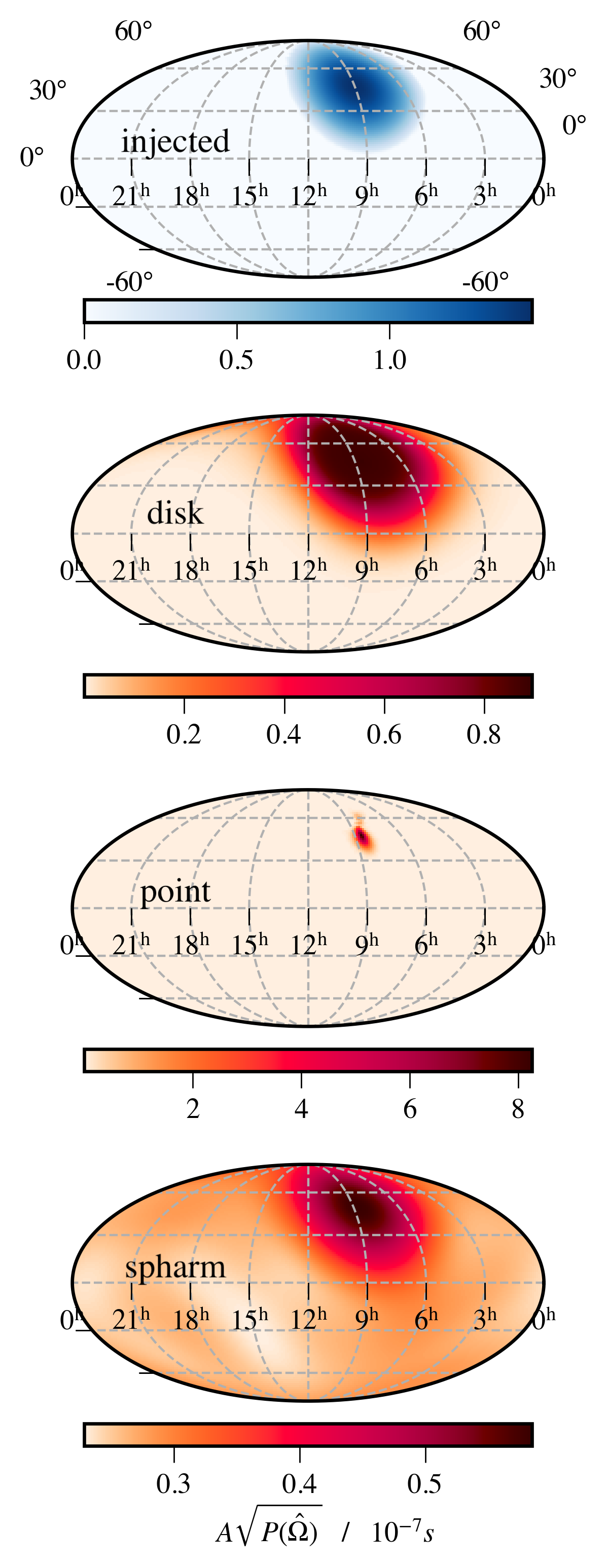}%
}
\subfloat[\label{fig:realpop_recongallery}]{%
  \includegraphics[width=0.492\textwidth]{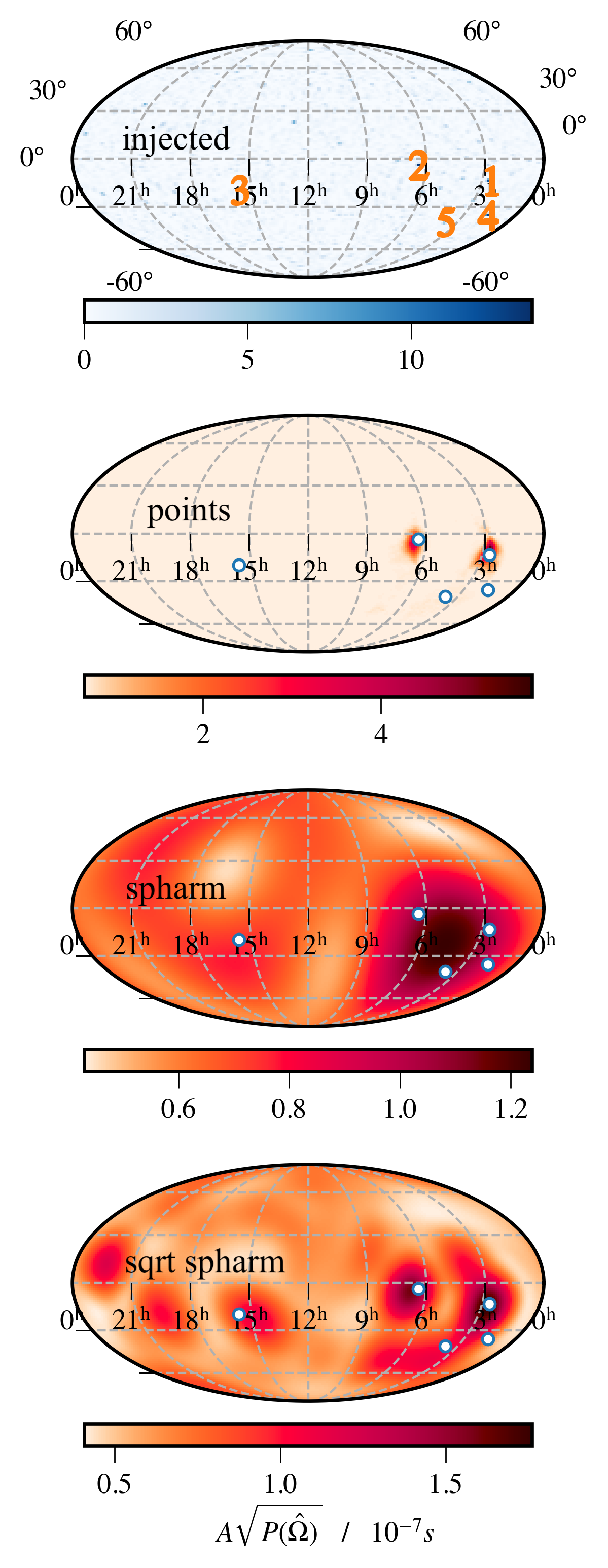}%
}
\caption{Posterior averaged skymaps for different parametrized mapping models tested against (a) a hotspot sky, and (b) a realistic SMBBH sky. The injected maps are shown as the top two panels in blue, with all reconstructions shown in red. For the SMBBH sky, the top 5 ranked loudest GW binaries are shown as numbered locations in the injected map and blue-edge white circles in the recovered maps.}
\end{figure*}

Finally, in the left column of Fig.~\ref{fig:hotspot_recongallery} we show the power reconstructions of all these different parametrizations of the hotspot sky, including the favored $l_\mathrm{max}=2$ spherical harmonic model. The injected map is shown at the top. The map reconstructions are averaged over the relevant model's posterior probability distribution. In all cases the region of hotspot power has clearly been identified and localized.

\begin{figure*}[htb]
\subfloat[\label{fig:realpop_sqrtspharm}]{%
  \includegraphics[width=0.492\textwidth]{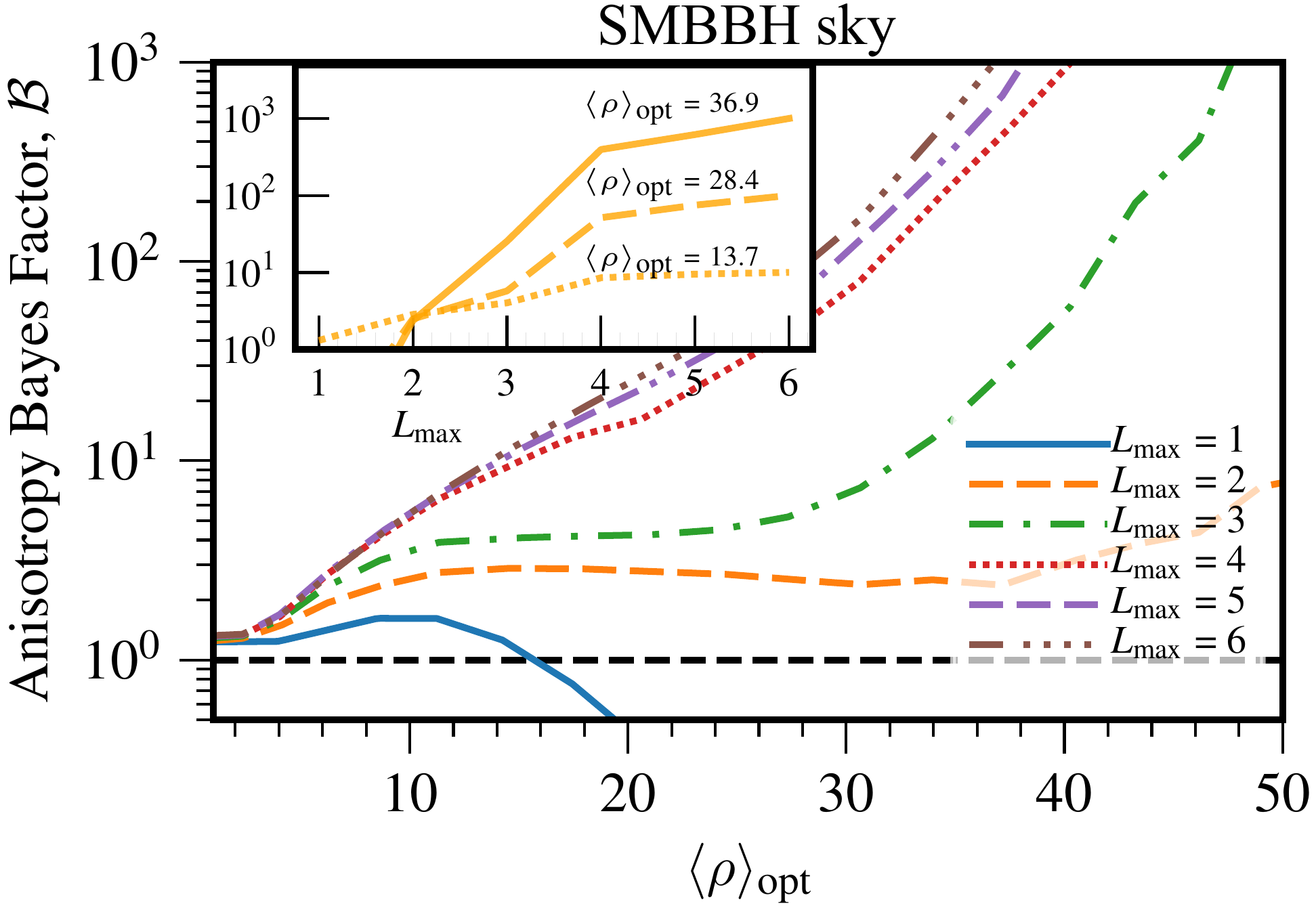}%
}
\subfloat[\label{fig:realpop_npoints}]{%
  \includegraphics[width=0.492\textwidth]{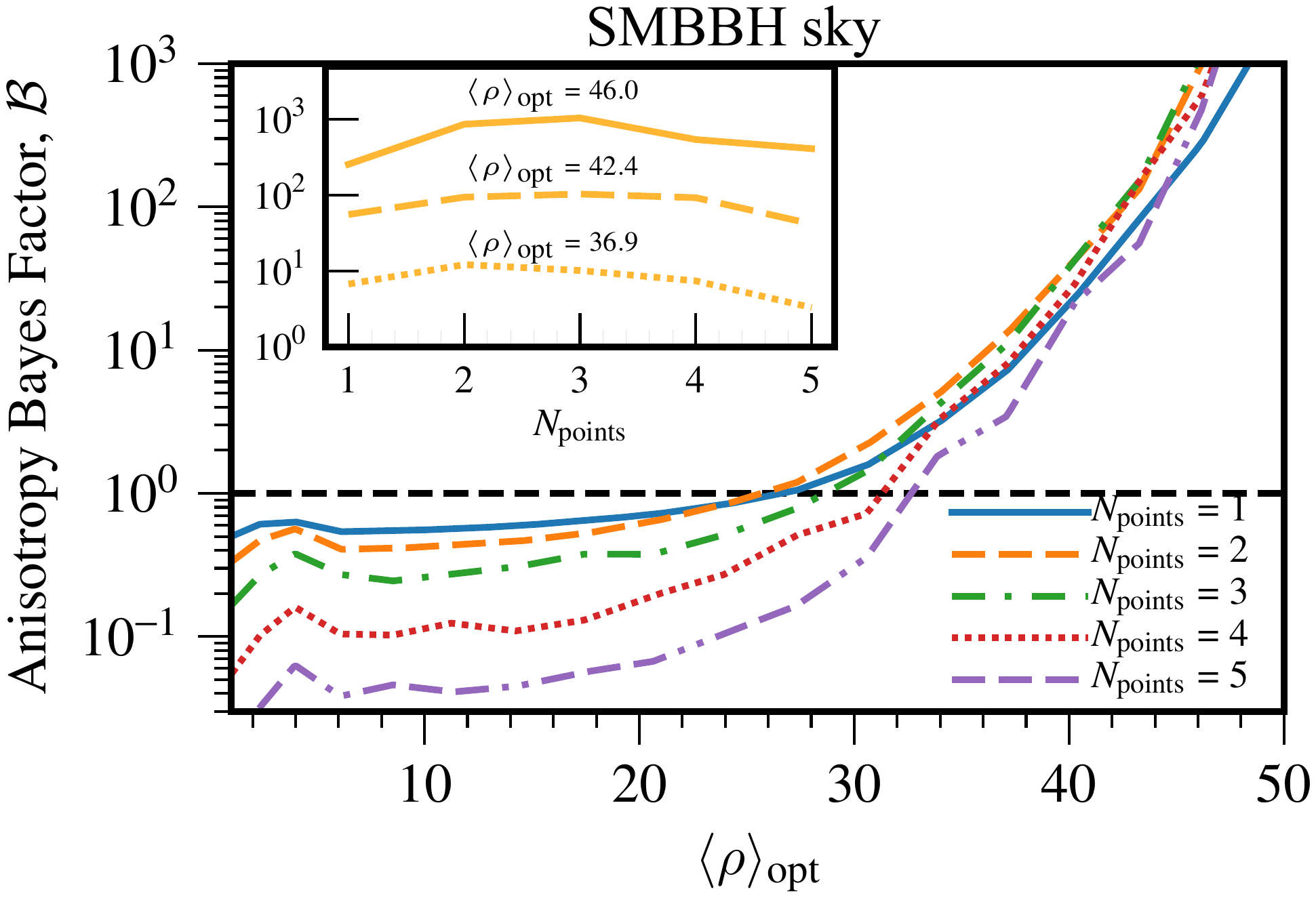}%
}
\caption{Comparison of anisotropy-vs-isotropy Bayes factor growth with SNR in a SMBBH sky for different choices of (a) $L_\mathrm{max}$ in spherical-harmonic modeling of $P(\hat\Omega)^{1/2}$, and (b) $N_\mathrm{points}$ in a multiple point-source model. Inset figures show vertical slices through the curves at different SNR values.}
\end{figure*}

\subsection{SMBBH sky tests}

A sky map produced by a population of SMBBHs is more challenging than the simple hotspot sky, as there is no obvious model that will a-priori be favored. The natural first test is between different $l_\mathrm{max}$ spherical harmonic recoveries, for which we show the Bayes factor growth versus SNR in Fig.~\autoref{fig:realpop_lmax_compare}. Note the SNR scale; arbitrating between anisotropic and isotropic power maps of SMBBHs may require very informative data. At low SNR the model with the fewest parameters is favored, corresponding to $l_\mathrm{max}=1$. As the data becomes more informative at SNRs ranging from $20-40$, an $l_\mathrm{max}=3$ model is favored. The inset figure hints at higher multipolar models becoming less disfavored at higher SNRs, where the data is better able to arbitrate deviations from isotropy and thus constrain smaller angular-scale features.  

We next reconstruct the sky by modeling the square root of the power map with spherical harmonics. The benefits of this approach with respect to straightforward spherical-harmonic power modeling were discussed in Sec.~\ref{sec:map_models}. The results are shown in Fig.~\autoref{fig:realpop_sqrtspharm}, where we see that $L_\mathrm{max}=1$ is dramatically disfavored; the squaring of the map to return to $P(\hat\Omega)$ produces features that are markedly different from isotropy and so are disfavored even at low SNR. However, increasing the $L_\mathrm{max}$ has a remarkable impact on the Bayes factor growth, with the data preferring ever larger $L_\mathrm{max}$ at all tested SNRs. The interpretation here is subtle, and related to how even a low-$l$ power anisotropy requires contributions from $L>l$ multipoles in the $P(\hat\Omega)^{1/2}$ model (see Appendix~\ref{sec:appendix}). Despite favoring ever higher $L_\mathrm{max}$, there is clearly no Occam penalty associated with the accompanying increase in the parameter volume. In fact, $L_\mathrm{max}=6$ reaches $\mathcal{B}=100$ at SNR $=28$, while $l_\mathrm{max}=3$ passes the same threshold later at SNR $=32$. 

Finally, we employ the most physically-motivated model for this kind of sky from the standpoint of its composition: point sources. The implementation differs from the previous hotspot-sky tests in two key ways: $(1)$ as mentioned in Sec.~\ref{sec:map_models}, we allow for multiple point sources across the sky, each with its own amplitude parameter; $(2)$ these point sources sit atop an isotropic GWB signal to emulate a realistic search scenario for multiple bright binaries resounding above a background. In Fig.~\ref{fig:realpop_npoints} we can see that any number of points is initially disfavored, with the data preferring an isotropic model below SNR $\sim25$. It is clear that the data lacks enough information to discriminate individual sources within the background at low SNR. Given that these point sources are being modeled on top of an isotropic signal, at low SNR the additional parameters of each source are simply incurring an Occam penalty. The penalty compounds for each additional modeled point source. It is only above SNR $\sim25$ that the data can discriminate individual sources, and by SNR $\sim35$ even preferring an Iso+$5$pts model. However, this model still performs worse than the root-power spherical-harmonic model, with Iso+$2$pts reaching $\mathcal{B}=100$ at SNR $=42$. Note that all of these SNR values at which various odds ratio thresholds are surpassed are specific to this SMBBH population realization. Hence while they are instructive and indicative of general trends, the SNR values at which anisotropy is favored may in fact be much lower if a small number of binaries swamp the entire GW sky at a given frequency. 

The posterior-averaged power reconstructions of all these different parametrizations of the SMBBH population are shown in Fig.~\ref{fig:realpop_recongallery}. The injected distribution is shown as the top map. The $5$ binaries that are loudest in GWs shown as numbered locations in the injected map, and blue-edged white circles in the reconstructed maps. Each parametrization localizes important regions of power to varying degrees, with the point-source and sqrt-spharm models best able to discriminate individual sources. In the latter case this is not obvious from the parameter posteriors, requiring map recovery to really visualize the strength of this model.

\section{Conclusions}
\label{sec:conclusions}

We have presented a general-purpose scheme to rapidly compute the inter-pulsar correlation signatures of anisotropic GW skies. This scheme is intended to be fast and flexible for embedding within a Bayesian analysis pipeline, and we have studied this use case here. We considered several different parametrized representations of anisotropic power; namely $(i)$ a spherical-harmonic power decomposition, $(ii)$ a spherical-harmonic decomposition of the square-root power, $(iii)$ lighting up multiple pixels atop an isotropic background, $(iv)$ a disk of power to model an extended region of GW emission. Our pipeline was tested against two very different power distributions, corresponding to an obvious region of excess power (a ``hotspot''), and a realistic GW sky generated by a population of SMBBHs. A key component of our Bayesian modeling was model selection, to not only determine when the data favored anisotropy over isotropy, but also to optimize various model hyper-parameters, e.g. the maximum multipole of spherical-harmonic modeling, and the number of pixels that were lit up to represent single sources.

For the hotspot sky, we found that the most apt model from a Bayesian perspective was a disk of power, with even a single point source being favored over spherical-harmonic models with $l_\mathrm{max}<6$. This happens because Bayesian model selection will favor the most parsimonious model that can represent the information in the data: a disk or single pixel can do so here with only a few parameters, whereas spherical harmonics require $(l_\mathrm{max}+1)^2$ parameters. For the SMBBH sky, our analysis marks the first time that robust Bayesian PTA techniques have been brought to bear on such a realistic GW sky. We found that the model that performed best was the spherical-harmonic modeling of square-root of the power distribution, which avoided the restrictive prior condition of the usual spherical-harmonic power modeling, and which did not seem to suffer any Occam penalties through the inclusion of higher multipoles. Nevertheless, the model that lit up multiple pixels atop an istropic GW background provides the most physically-motivated and easily-interpreted representation of a SMBBH population, and did in fact perform better than both the $l_\mathrm{max}\leq 2$ and $L_\mathrm{max}\leq 2$ models.

Constraining deviations from isotropy in the nanohertz GWB is an enticing goal for the era beyond initial PTA detection. It is possible that as the GW signal significance grows in our data, loud binaries will first emerge from the background ensemble in excess angular power searches, before eventually having their binary model's constrained. As such, the fast and flexible techniques presented here form an important link in the chain of scientific milestones for PTAs, helping to kick-start single-source localization efforts even before other binary properties are known. 

There are a variety of potential extensions to this work that build from assumptions we have made. 

\begin{enumerate}
    \item We have implicitly assumed a model for the GW angular power distribution that is the same at all frequencies. But this will not necessarily be the case in reality. In fact, the coalescence timescale of SMBBHs in the PTA band can be $\sim$Myrs, implying almost negligible evolution of the system within our observational window. For all intents and purposes, these are monochromatic signals. Thus one or several bright binaries may dominate over others in one frequency bin, but not in another, giving variable power maps across frequency. Indeed, as one moves toward higher frequencies in the PTA band, sources become sparser, implying a greater prospect for individual-source and anisotropic signatures. Adapting our framework for this case is trivial, simply requiring frequency-indexed overlap reduction functions.
    \item The SMBBH skymap was generated using the inclination-averaged strain of each binary, for which we then assumed that the power maps in $+$ and $\times$ polarizations were equal. However, differences in these power polarization maps encode information, and any imbalance may indicate the presence of a bright binary dominating in a certain pixel. Relaxing our assumption of equal power in each polarization may further help in recognizing bright single sources, and can be easily achieved within our framework by allowing the polarization maps to be independently modeled.  
    \item We have considered a statistically representative SMBBH population in order to produce a realistic GW skymap, however different populations will produce variations in anisotropy. Our population has several bright (somewhat clustered) sources that we were able to localize, but we did not consider a population where one binary overwhelmingly dominated the entire strain budget (which can sometimes happen). From a single-source detection standpoint, the population we considered may seem somewhat pessimistic, but this prospect varies significantly from realization to realization. In future work we will assess a range of population realizations to consider how well different models perform as a function of the power concentration in the loudest sources. It is likely that the single pixel modeling will be the most preferred model in a realization where a single source really sticks out.
\end{enumerate}

In the spirit of open science, our code to rapidly compute PTA overlap reduction functions from a given angular power distribution is available at \href{https://github.com/stevertaylor/gworf}{https://github.com/stevertaylor/gworf}.

\begin{acknowledgments}

We thank Sharan Banagiri, Vuk Mandic, Joe Romano, Siyuan Chen, Bence Bécsy, Eric Thrane, Ethan Payne, and Marc Kamionkowski for fruitful discussions. We are also indebted to our colleagues in the NANOGrav Collaboration and the International Pulsar Timing Array for useful discussions. Much of this work was performed at the Jet Propulsion Laboratory, where SRT was supported by appointment to the NASA Postdoctoral Program, administered by Oak Ridge Associated Universities and the Universities Space Research Association through a contract with NASA. SRT was also supported by the NANOGrav NSF Physics Frontier Center award number 1430284. This work was supported in part by National Science Foundation Grant No. PHYS-1066293 and by the hospitality of the Aspen Center for Physics. A.S. is supported by the European Research Council (ERC) under the European Union’s Horizon 2020 research and innovation program ERC-2018-COG under grant
agreement No 818691 (B Massive). A majority of the computational work was performed on the Nemo cluster at UWM supported by NSF grant No. 0923409. Some of the results in this paper have been derived using the HEALPix (K.M. Górski et al., 2005, ApJ, 622, p759) package.
\end{acknowledgments}

\appendix
\label{sec:appendix}

\section{Decomposing power coefficients $c_{lm}$ in terms of square-root power coefficients $a_{LM}$}\label{sec:appA}
The calculation below requires Wigner-$3j$ identities which are most easily expressed with complex spherical harmonics. The real spherical harmonics are merely a linear combination of these complex spherical harmonics, so the qualitative results are the same in both.

For the power decomposition in spherical harmonics we have:
 \begin{equation}
        P(\hat{\Omega}) = \sum_{l=0}^{\infty}\sum_{m=-l}^{l} c_{lm} Y^m_l(\hat{\Omega}),
        \label{eq:app_power}
    \end{equation}
while for the square-root power decomposition in spherical harmonics we have:
\begin{equation}
    P(\hat{\Omega})^{1/2} = \sum_{L=0}^{\infty}\sum_{M=-L}^{L} a_{LM} Y^M_L(\hat{\Omega}),
    \label{eq:app_sqrtpower}
\end{equation}
where the notation $Y^m_l, Y^M_L$ indicates the complex spherical harmonics.
  
\begin{figure*}[htb]
\subfloat[\label{fig:app_power}]{%
  \includegraphics[width=0.325\textwidth]{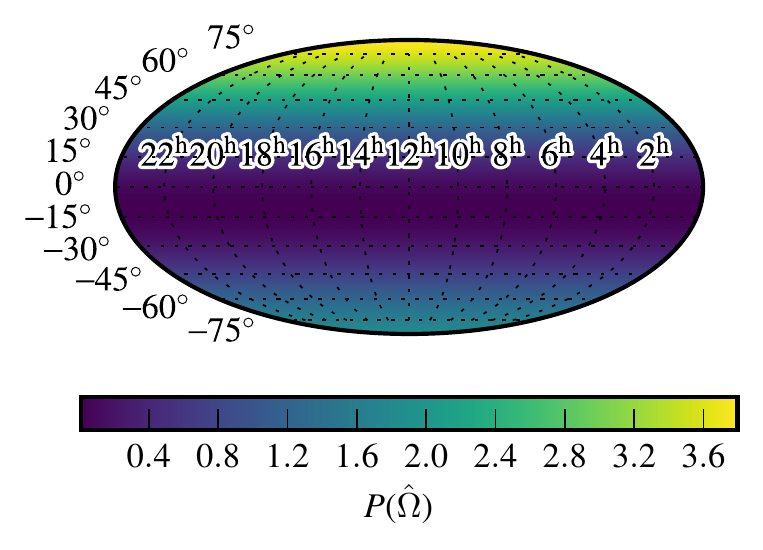}%
}
\subfloat[\label{fig:power_rootpower_spectra}]{%
  \includegraphics[width=0.325\textwidth]{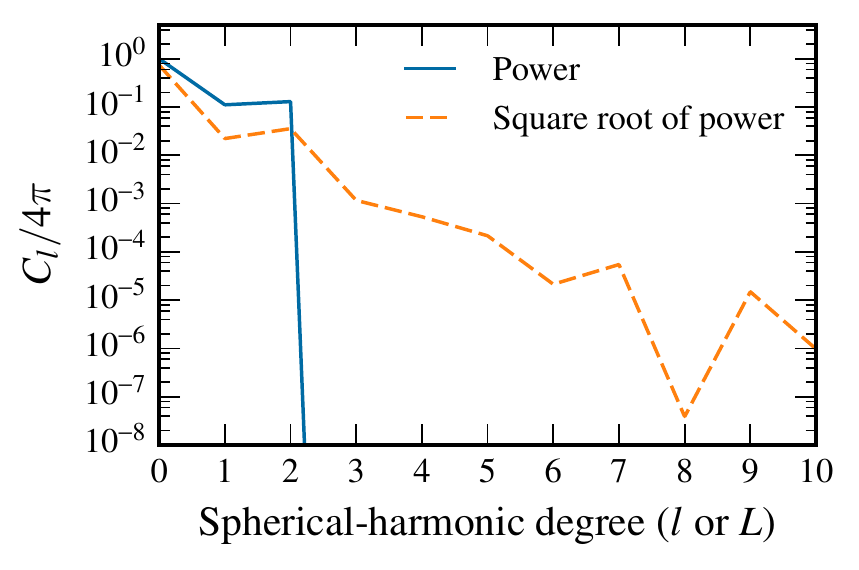}%
}
\subfloat[\label{fig:rootpower_match}]{%
  \includegraphics[width=0.325\textwidth]{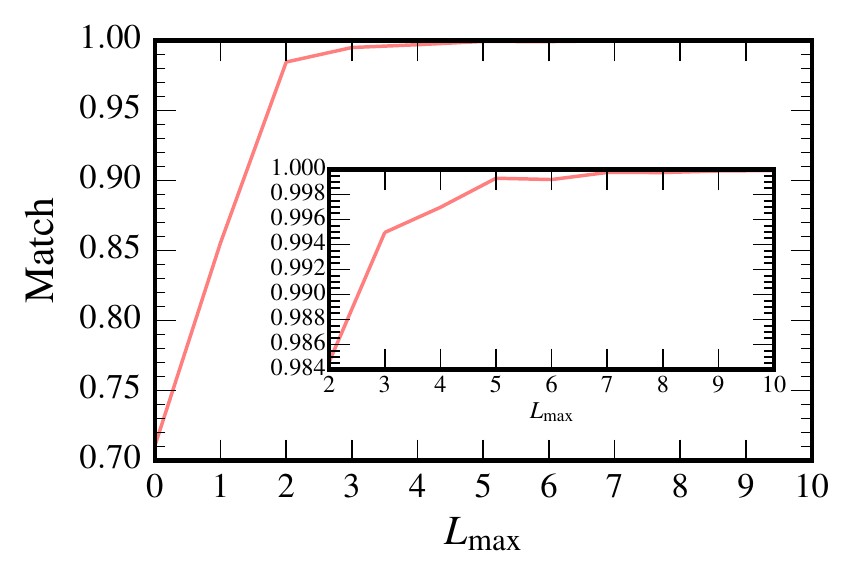}%
}
\caption{(a) the tested power map with monopole, dipole, and quadrupole anisotropies. It is fully described by $P(\hat\Omega)=1+\cos\theta+0.9\times(3\cos^2\theta-1)$. (b) the angular power spectra of $P(\hat\Omega)$ and $P(\hat\Omega)^{1/2}$. (c) the match between the original power map and a reconstruction from the squaring of a restricted $L$ decomposition of $P(\hat\Omega)^{1/2}$.}
\label{figure:power_rootpower_decomposition}
\end{figure*}

Squaring \autoref{eq:app_sqrtpower} and equating to \autoref{eq:app_power} gives:
\begin{align}
P(\hat\Omega) &= \sum_{LM}\sum_{L'M'} a_{LM}a_{L'M'} Y^M_L(\hat\Omega)Y^{M'}_{L'}(\hat\Omega) \nonumber\\
&= \sum_{lm} c_{lm} Y^m_l(\hat{\Omega}).
\end{align}
Thus, solving for $c_{lm}$ gives:
\begin{align}
&c_{lm} =  \sum_{LM}\sum_{L'M'} a_{LM}a_{L'M'} \int_{S^2} \mathrm{d}\hat\Omega\; Y^{m*}_l(\hat\Omega) Y^M_L(\hat\Omega) Y^{M'}_{L'}(\hat\Omega) \nonumber\\
&= (-1)^m \sum_{LM}\sum_{L'M'} a_{LM}a_{L'M'} \int_{S^2} \mathrm{d}\hat\Omega\; Y^{m}_l(\hat\Omega) Y^M_L(\hat\Omega) Y^{M'}_{L'}(\hat\Omega)
\label{eq:clm_sumalm}
\end{align}
where the triple spherical-harmonic integral in \autoref{eq:clm_sumalm} can be expressed as a product of Wigner-$3j$ symbols with the following identity:
\begin{align}
&\int_{S^2} \mathrm{d}\hat\Omega\; Y^{m}_l(\hat\Omega) Y^M_L(\hat\Omega) Y^{M'}_{L'}(\hat\Omega) = \nonumber\\
&\sqrt{\frac{(2l+1)(2L+1)(2L'+1)}{4\pi}}\begin{pmatrix} 
l & L & L' \\ -m & M & M'\end{pmatrix}\begin{pmatrix} l & L & L' \\ 0 & 0 & 0\end{pmatrix}.
\end{align}
Hence, the power coefficients $c_{lm}$ can be written as the following (infinite) summation over $a_{LM}$ terms:
\begin{align} \label{eq:clm_from_alm}
c_{lm} &= (-1)^m \sum_{L=0}^{\infty}\sum_{M=-L}^{L}\sum_{L'=0}^{\infty}\sum_{M'=-L}^{L'} a_{LM}a_{L'M'} \times \nonumber\\
&\sqrt{\frac{(2l+1)(2L+1)(2L'+1)}{4\pi}} 
\begin{pmatrix} l & L & L' \\ -m & M & M'\end{pmatrix}\begin{pmatrix} l & L & L' \\ 0 & 0 & 0\end{pmatrix}.
\end{align}
\autoref{eq:clm_from_alm} shows that even a low order coefficient of power anisotropy may need contributions from higher orders in the square root of power, such that $L_\mathrm{max}>l_\mathrm{max}$. We investigate this with a numerical example, where we create an $N_\mathrm{side}=256$ HEALPix map of GWB power (shown in Fig.\ \ref{fig:app_power}) containing anisotropy up to and including $l=2$, such that $P(\hat\Omega) = 1+ \cos\theta + 0.9\times(3\cos^2\theta-1)$. We also take the square root of this map, and use HEALPix functions to compute the angular power spectra of both $P(\hat\Omega)$ and $P(\hat\Omega)^{1/2}$. These spectra are shown in Fig.\ \ref{fig:power_rootpower_spectra}, where we see that $P(\hat\Omega)$ contains anisotropic structure up to and including $l=2$, but higher degrees drop to zero (within numerical precision). On the other hand, $P(\hat\Omega)^{1/2}$ has anisotropic structure in degrees beyond quadrupole ($L>2$) which are needed for a complete reconstruction of $P(\hat\Omega)$. However, in practice we can achieve an adequate approximation of $P(\hat\Omega)$ by squaring $P(\hat\Omega)^{1/2}$ with $L\leq 2$. This is shown in Fig.\ \ref{fig:rootpower_match}, where we see the normalized match (inner product) between the original power map and a reconstruction from the squaring of a restricted $L$ decomposition of $P(\hat\Omega)^{1/2}$. For $L_\mathrm{max}=2$ the match is better than $98\%$.

\section{Product space sampling}\label{sec:appB}
This technique recasts Bayesian model selection as a parameter estimation problem \citep{cc95,godsill01,hee16,abb+18b}. We define a hypermodel, $\mathcal{H}$, whose parameter space is the Cartesian product of the individual spaces of all sub-models under consideration, along with an additional model indexing parameter, $n$. This model indexing parameter is discrete, but continuous sampling methods can simply be applied to a variable that is cast to an integer in the likelihood, or for which we define behavior based on parameter ranges. 

The practical implementation of this is rather straightforward. At a given iteration in the sampling process we cast the model indexing parameter to an integer, which then indicates the ``active'' sub-model  for the likelihood evaluation. The hypermodel parameter space, $\mathbf{\theta_*}$, is partitioned into the sub-model spaces, and the corresponding parameters of the active sub-model are passed to the relevant likelihood function. Thus the parameters of the inactive sub-models do not contribute to, and are not constrained by, the active likelihood function. As sampling proceeds, the model indexing parameter will switch between all sub-models to the end result that the relative fraction of sampling iterations spent in each sub-model provides an estimate of the posterior odds ratio. 

This result has been derived in other places (see e.g. \citep{hee16}), but we reiterate it here for completeness. Consider constructing the marginalized posterior distribution of the model indexing parameter from the output of MCMC sampling:
\begin{align}
    p(n|\mathbf{d},\mathcal{H_*}) &= \int p(\mathbf{\theta_*},n|\mathbf{d},\mathcal{H_*})d\mathbf{\theta} \nonumber\\
    &= \frac{1}{\mathcal{Z_*}} \int p(\mathbf{d}|\mathbf{\theta_*},n,\mathcal{H_*})p(\mathbf{\theta_*},n|\mathcal{H_*})d\mathbf{\theta}
\end{align}
where $\mathcal{Z_*}$ is the hypermodel evidence. For a given $n$ the hypermodel parameter space is partitioned into active, $\mathbf{\theta_n}$, and inactive, $\mathbf{\theta_{\bar{n}}}$, parameters, $\theta_* = \{\theta_n,\theta_{\bar{n}}\}$, where the likelihood $p(\mathbf{d}|\mathbf{\theta_*},n,\mathcal{H_*})$ is independent of the inactive parameters. The prior term can then be factorized:
\begin{equation}
    p(\mathbf{\theta_*},n|\mathcal{H_*}) = p(\mathbf{\theta_n}|\mathcal{H}_n)p(\mathbf{\theta_{\bar{n}}}|\mathcal{H}_{\bar{n}})p(n|\mathcal{H_*})
\end{equation}
such that
\begin{align}
    p(n|\mathbf{d},\mathcal{H_*}) &= \frac{p(n|\mathcal{H_*})}{\mathcal{Z_*}} \int p(\mathbf{d}|\mathbf{\theta_n},\mathcal{H}_n) p(\mathbf{\theta_n}|\mathcal{H}_n) d\mathbf{\theta_n} \nonumber\\
    &= \frac{p(n|\mathcal{H_*})}{\mathcal{Z_*}} \mathcal{Z}_n
\end{align}
where $\mathcal{Z}_n$ is the evidence for sub-model $n$, and we have implicitly marginalized over inactive parameters since they only appear in the prior term $p(\mathbf{\theta}_{\bar{n}}|\mathcal{H}_{\bar{n}})$ which simply integrates to unity. Thus the posterior odds ratios between two models is given by:
\begin{equation}
    \mathcal{P}_{12} = \frac{p(n_1|\mathcal{H_*})\mathcal{Z}_1}{p(n_2|\mathcal{H_*})\mathcal{Z}_2} = \frac{p(n_1|\mathbf{d},\mathcal{H_*})}{p(n_2|\mathbf{d},\mathcal{H_*})}
\end{equation}
where the hypermodel evidence cancels in this ratio of the two sub-models. We add an additional refinement to this method, in that we perform a pilot run designed to provide an initial estimate of the posterior odds ratio, $\tilde{\mathcal{P}}_{12}$ then follow this up with a focused run for an improved estimate. In the focused run, we weight model $1$ by $1 / (1 + \tilde{\mathcal{P}}_{12})$ and model $2$ by $\tilde{\mathcal{P}}_{12} / (1 + \tilde{\mathcal{P}}_{12})$. This has the benefit of improving mixing between models when the data significantly favor one model over another. The posterior odds ratio from the focused run is then re-weighted to provide an improved estimate of $\mathcal{P}_{12}$. We estimate the uncertainty using methods for Reversible Jump MCMC \citep{cl15}, as the approaches are analogous. This product-space estimator of the posterior odds ratio is simple to implement, applicable to high dimensional parameter spaces, and allows direct model comparison.

\bibliography{apjjabb,bib}

\end{document}